\documentclass[aps,prd,10pt,twocolumn,a4paper,superscriptaddress,%
               nofootinbib,floatfix,longbibliography]{revtex4-2}

\usepackage{amsmath,amssymb,amsfonts}
\usepackage{graphicx}
\usepackage{tikz}
\usetikzlibrary{arrows.meta}
\usepackage{bm}
\usepackage{xcolor}
\usepackage{enumitem}

\usepackage[colorlinks = true,
            linkcolor  = blue,
            urlcolor   = blue,
            citecolor  = blue,
            anchorcolor= blue]{hyperref}

\newcommand{\tc}{\tilde{c}}
\newcommand{\tceff}{\tilde{c}_{\rm eff}}
\newcommand{\Zt}{\mathbb{Z}_2}
\newcommand{\Uone}{U(1)}
\newcommand{\ld}{\ell_d}
\newcommand{\vbar}{\bar{v}}
\newcommand{\Amp}{\mathcal{A}}
\newcommand{\Aobs}{\mathcal{A}_{\rm obs}}
\newcommand{\knr}{k_{\rm nr}}
\newcommand{\lam}{\lambda}
\newcommand{\kap}{\kappa}
\newcommand{\CTY}{CTY}

\begin{document}

\title{The friction-era VOS amplitude of a
       \texorpdfstring{$\Zt$}{Z2} string network
       from nematic disclination data}

\author{Dimitrios Efstratiou}
\email{d.efstratiou@uoi.gr}
\affiliation{Department of Physics, University of Ioannina, GR-45110, Ioannina, Greece}
\author{Evangelos Achilleas Paraskevas}
\email{e.paraskevas@uoi.gr}
\affiliation{Department of Physics, University of Ioannina, GR-45110, Ioannina, Greece}
\author{Leandros Perivolaropoulos}
\email{leandros@uoi.gr}
\affiliation{Department of Physics, University of Ioannina, GR-45110, Ioannina, Greece}

\date{\today}

\begin{abstract}
A tangle of line defects coarsens as the mean spacing $L$ between neighbouring lines grows. In a viscous medium the motion is overdamped, and the velocity-dependent one-scale (VOS) model predicts a late-time attractor $L^{2}=\Amp\,\ld\,t$, with $\ld=T/\Gamma$ the ratio of line tension to drag and $\Amp$ a dimensionless amplitude. The growth law $L\propto t^{1/2}$ holds for every value of the three model parameters---the momentum parameter $k\le1$, the sink coefficient $\tc$, and the curvature ratio $\lam\equiv R/L$---since all three enter only through $\Amp=\kap(\kap+\tc)$, $\kap\equiv k/\lam$. The exponent therefore constrains none of them, and the amplitude is the only quantity a density history can deliver. We measure it from the disclination data of Chuang, Turok and Yurke on a nematic liquid crystal, whose companion measurement of loop collapse fixes $\ld$ on the same samples, so that the material constants of 5CB cancel between the two. Treating the unmatched per-quench $\ld$ as a nuisance parameter with a Gaussian prior and marginalizing it analytically, we obtain $\Amp=10.0^{+1.3}_{-1.1}$ from the three quenches in the $234~\mu$m cell; the fourth, the only one in the thinner $158~\mu$m cell, gives $\Amp=3.0^{+0.8}_{-0.6}$ and is fitted separately. Converting either into $\tc$ requires $\lam$, which these data do not determine, so the result is a curve, $\tc(\lam)=\Amp\lam/k-k/\lam$, not a number. At $\lam=1$ with $k\le1$ they give $\tc\ge9.0$ and $\tc\ge2.0$, against $\tc=0.23$--$0.57$ from relativistic $\Uone$ simulations, values reached only at $\lam\simeq0.33$--$0.35$ and $0.62$--$0.68$. We know of no VOS calibration for a cosmological $\Zt$ network, so we cannot attribute the excess to topology, and we list what a repeat experiment would have to measure.
\end{abstract}

\maketitle

\section{Introduction}
\label{sec:intro}

When a physical system is cooled or compressed rapidly through a symmetry-breaking phase transition, it orders locally before it can order globally. Regions that are too far apart to communicate make independent choices of the new ground state, and where those choices are incompatible the system is left with defects it cannot remove. This is the Kibble mechanism. If the space of available ground states---the vacuum manifold---contains loops that cannot be shrunk to a point, the defects left behind are lines. The same argument applies to the early universe and to a laboratory quench, and produces line defects in both~\cite{Kibble1976,Kibble1980,VilenkinShellard,KlemanLavrentovich,Zurek1985,Zurek1996}.

This paper is about how such a tangle of lines coarsens with time, and about which property of that coarsening a laboratory experiment can actually measure. We use the word \emph{coarsening} throughout to mean the growth of the characteristic scale of the tangle, equivalently the decay of the defect density.

\subsection{The physical picture}
\label{sec:picture}

A rapid quench leaves a dense, disordered tangle of line defects. The tangle is described by one characteristic scale $L$, the typical distance between neighbouring lines. As time passes $L$ grows and the density of lines falls. We are concerned with the case where the lines move through a viscous medium. In that case their motion is \emph{overdamped}: inertia is negligible compared with drag, so nothing moves except under its own curvature. A curved segment is pulled toward its centre of curvature by its own tension, and drifts at whatever speed makes the drag balance that pull. It does not accelerate, and it stops as soon as it is straight.
\begin{figure}[t]
  \centering
  \begin{tikzpicture}[>=Stealth,scale=0.78]
    \begin{scope}
      \draw[very thick] (0,0) circle (0.92);
      \draw[gray,dashed] (0,0) circle (0.55);
      \draw[gray,dashed] (0,0) circle (0.22);
      \foreach \a in {45,135,225,315}{\draw[->,thick] (\a:1.30) -- (\a:1.00);}
      \node at (0,-1.80){\footnotesize (a) curve shortening};
      \node at (0,-2.30){\footnotesize $r^{2}=2\ld(t_{0}-t)$};
    \end{scope}
    \begin{scope}[shift={(3.6,0)}]
      \draw[very thick] (-0.95,0.95) .. controls (-0.10,0.45) and (-0.10,-0.45) .. (-0.95,-0.95);
      \draw[very thick] ( 0.95,0.95) .. controls ( 0.10,0.45) and ( 0.10,-0.45) .. ( 0.95,-0.95);
      \draw[->,thick] (-0.62,0) -- (-0.30,0);
      \draw[->,thick] ( 0.62,0) -- ( 0.30,0);
      \node at (0,-1.25){\footnotesize before};
    \end{scope}
    \draw[-{Stealth[length=2.6mm]},thick] (4.95,0) -- (5.65,0);
    \begin{scope}[shift={(7.0,0)}]
      \draw[very thick] (-0.95,0.95) .. controls (-0.35,0.30) and (0.35,0.30) .. (0.95,0.95);
      \draw[very thick] (-0.95,-0.95) .. controls (-0.35,-0.30) and (0.35,-0.30) .. (0.95,-0.95);
      \node at (0,-1.25){\footnotesize after};
    \end{scope}
    \node at (5.3,-1.80){\footnotesize (b) removal at contact};
    \node at (5.3,-2.30){\footnotesize $\simeq\tc L$ removed};
  \end{tikzpicture}
  \caption{The two ways a tangle loses length, and the reason the coarsening exponent cannot tell them apart. (a) Curve shortening: a curved segment drifts toward its centre of curvature and is consumed against the viscous medium. No topology changes, and for an isolated loop this is the whole of the dynamics, which is why the loop-collapse measurement of Ref.~\cite{Chuang1991PRL} isolates the transport coefficient $\ld$ cleanly. (b) Removal at contact: two segments meet and reconnect, and a portion of the network ceases to belong to it. The sink coefficient $\tc$ measures the length removed per encounter in units of $L$; it is a length ratio and not a probability, and is therefore not bounded by unity. Both processes are driven by the same curvature-induced  drift, so both give $L\propto t^{1/2}$ and differ only in the prefactor
  $\Amp$.}
  \label{fig:channels}
\end{figure}
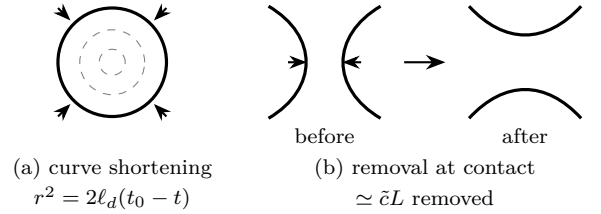
Length then disappears in two ways, and the distinction between them is the subject of this paper. Figure~\ref{fig:channels} shows both. The first way is \emph{curve shortening}. A drifting segment dissipates energy against the viscous medium. Since its energy is proportional to its length, the length itself is consumed. This involves no change of topology. It is the entire dynamics of a single isolated loop of defect line, which simply shrinks and disappears (which is why a measurement of loop collapse isolates the transport properties of the medium cleanly, with no network physics mixed in). The second way is \emph{contact}. Segments that meet reconnect or annihilate, and a portion of the network ceases to be part of it. The coefficient measuring the length removed per such encounter, in units of $L$, is the sink coefficient $\tc$. 

Both routes are driven by the same curvature-induced drift, so both produce the same square-root growth of $L$ with time. The coarsening exponent therefore distinguishes neither of them. They differ only in how fast that growth proceeds, that is, in the prefactor. That prefactor is the amplitude $\Amp$ of the abstract and is what we are going to infer: it is the factor by which the tangle coarsens faster than a collection of freely collapsing loops of its own curvature radius would.

\subsection{The VOS model and its calibration}
\label{sec:vosintro}

The standard analytic description of late-time coarsening is the velocity-dependent one-scale (VOS) model of Martins and Shellard~\cite{MartinsShellard1996,PhysRevD.54.2535,PhysRevB.56.10892,MartinsShellard2002,Martins:1997nb,MartinsMooreShellard2004,Martins2016}. Rather than following every segment of line, it averages the microscopic dynamics down to two coupled ordinary differential equations, one for the characteristic scale $L(t)$ and one for the root-mean-square velocity $\vbar(t)$ of the lines. That reduction is not free: closing the two equations requires two phenomenological numbers that the averaging cannot supply. These are the curvature--momentum parameter $k$ and the sink coefficient $\tc$ introduced above, and they must be calibrated against simulations or experiment.

Both have so far been calibrated only for relativistic, frictionless, oriented $\Uone$ cosmic strings---that is, for strings moving freely at near light speed rather than being dragged through a medium, and carrying a direction, so that a string and an anti-string are distinguishable. Even there the numbers are regime dependent. Simulations of the Abelian--Higgs model, in which the string core is resolved as a configuration of fields, and flat-space Nambu--Goto simulations, in which the string is idealised as an infinitely thin line with tension only, give $\tc\simeq0.57$; expanding radiation-era Nambu--Goto simulations give $\tc\simeq0.23$. Reference~\cite{Martins2016} reads these two values as the bare and renormalized \emph{chopping efficiency}, the rate at which the long-string network is chopped into loops~\cite{MooreMartinsShellard2002,MartinsMooreShellard2004,Vanchurin2005}. These two numbers anchor current calculations of the stochastic gravitational-wave background from cosmic strings~\cite{BlancoPilladoOlumShlaer2014,Auclair2020,NANOGrav2023}. More recently, high-resolution field theory simulations have extended the VOS model to account explicitly for the energy carried away by scalar and gauge radiation~\cite{Correia:2019bdl,Correia:2021tok}; by separating that channel from loop production, the modern calibration revises the $\Uone$ Abelian--Higgs value downward to $\tc\approx0.30$~\cite{Correia:2021tok}.

None of these calibrations covers the \emph{friction era}: the regime in which defects are dragged through a medium, so that their velocity is set by a balance between tension and drag rather than by acceleration. In cosmology this is the early period in which strings scatter off the ambient plasma; in a liquid crystal it is the whole of the accessible dynamics, since a disclination line is always immersed in a viscous fluid. It is the regime in which the amplitude measured here applies.

\subsection{Why a liquid crystal is relevant} \label{sec:whylc}

A nematic liquid crystal is a fluid of rod-like molecules that share a common average orientation, described by a unit vector field $\bm n$ called the director, but have no positional order. Since the molecules have no head--tail distinction, $\bm n$ and $-\bm n$ describe the same state, and the director is defined only up to that identification. The line defects of a nematic are called \emph{disclinations}: lines about which the director fails to return to itself, returning instead to an orientation equivalent to the original.

Which line defects a medium admits is fixed by the topology of its vacuum manifold $\mathcal{M}_0$ (the order parameter, or degeneracy, space in condensed-matter texts~\cite{KlemanLavrentovich}), the set of order-parameter values that leave the thermodynamic potentials unchanged. The relevant classification is the first homotopy group $\pi_1(\mathcal{M}_0)$, which sorts closed loops in that space into classes that can be deformed into one another; a non-trivial class corresponds to a line defect that cannot be removed by any local rearrangement. The identification above gives $\mathcal{M}_0=\mathbb{R}P^2=S^2/\{\bm n\sim-\bm n\}$, for which $\pi_1(\mathbb{R}P^2)\cong\Zt$: one non-trivial class and one trivial class. Type-$\frac{1}{2}$ disclinations correspond to the non-trivial class, the loops running between antipodal points of $\mathcal{M}_0$; these cannot be contracted, which endows the lines with singular cores and high energy per unit length. Type-1 disclinations correspond to the trivial class, whose loops close on themselves and can be contracted; the director then escapes into the third dimension and removes the core, rendering these lines non-singular and energetically subdominant~\cite{KlemanLavrentovich,nakahara2003geometry,Chuang1991PRL}. The same manifold also supports point defects (classified by $\pi_2$) and textures ($\pi_3$). Crucially, $+\frac{1}{2}$ and $-\frac{1}{2}$ lines belong to the same $\Zt$ class and can be deformed into one another, so a nematic disclination is its own antidefect and two segments brought together annihilate whichever way round they meet. An oriented $\Uone$ network has $\pi_1\cong\mathbb{Z}$ and admits no such freedom: two like-signed strings cannot annihilate, only a string and an antistring~\cite{KlemanLavrentovich,ChuangPRE1993,McGraw:1997nx}.

The connection to cosmology is that the VOS equations do not know what made the lines: the momentum parameter $k$ and the sink coefficient $\tc$ are defined by the averaging, not by the microphysics, so the same two numbers close the equations for a nematic disclination tangle and for a cosmological string network. Their values need not agree, and in the friction era the two systems do not even share a growth law\cite{Martins2016}, since the cosmological damping length grows with time. What a laboratory quench can supply is a measurement of $\tc$ in a regime where it has never been measured, and for a $\Zt$ topology for which we are not aware of any cosmological calibration.

Two qualifications should be recorded at once, since both bear on how the results below may be read. First, the analysis assumes that the type-$\frac{1}{2}$ lines dominate and define a single characteristic scale $L$. They are more energetic than the type-1 lines and hence determine the late-time dynamics~\cite{Chuang1991PRL}, and textures appear to be very rare~\cite{Chuang1991PRL}. The approximation is a reasonable one, but it is empirical rather than derived: Ref.~\cite{ChuangPRE1993} notes that the energies of the different defect types scale differently with the characteristic length, so that a single-scale description is not expected in advance, and rests the approximation on the observed dominance of the type-$\frac{1}{2}$ defects. That same reference finds the type-$\frac{1}{2}$ density to scale as expected while the loop and monopole densities do not, and reports that one of its two most abundant events is the absorption of a type-1 line into a type-$\frac{1}{2}$. Second, the self-inverse $\Zt$ class removes a constraint that oriented networks impose, so an enhanced sink would be a reasonable expectation. But $\tc$ counts both ways the network loses length at contact, chopping into loops and direct annihilation, and no fit to the density can divide them. An elevated $\tc$ below is therefore consistent with an annihilation channel, but does not single it out: the same value is reached with the loop channel alone if the tangle is sufficiently wiggly. Reference~\cite{ChuangPRE1993} takes three-dimensional coarsening to be driven mainly by tension straightening curved segments rather than by string--string annihilation, though it states this as an expectation and does not test it.

\subsection{The dataset, and what it can deliver} \label{sec:dataintro}

Chuang, Turok and Yurke~(\CTY)~\cite{Chuang1991PRL,Chuang1991Science} subjected the nematic liquid crystal 5CB to a rapid pressure jump---a pressure quench---that carried it across the isotropic-to-nematic transition, and measured the resulting decay of the defect density, $\rho\propto t^{-1.02\pm0.09}$. Later work confirmed that scaling~\cite{BowickChandar1994,Pargellis1992,ChuangPRE1993,Bray1994}. We digitized their data and analyze them with the VOS model~\cite{Martins2016,Coelho:2026oeg} in its damped, non-relativistic limit, where rotational viscosity dominates inertia ($\dot{\vbar}=0$) and the network is locked onto the overdamped attractor~\cite{ImuraOkano,deGennesProst}.

By an \emph{attractor} we mean throughout a late-time solution of the evolution equations that is approached from a wide range of initial configurations, and which thereafter depends only on the parameters of the model and not on how the tangle was formed. As Sec.~\ref{sec:model} shows, the overdamped attractor of the VOS equations is
\begin{equation}
  L^{2}(t)=\Amp\,\ld\,t,
  \label{eq:attractor_preview}
\end{equation}
and the three parameters $k$, $\tc$ and $\lam$ enter it only through the single dimensionless number $\Amp=\kap(\kap+\tc)$. Here $\ld$ is the transport coefficient of the medium, with dimensions of area per unit time.

Two consequences follow immediately, and they set the scope of this paper. Since $\Amp$ and $\ld$ are constants, Eq.~\eqref{eq:attractor_preview} gives $L\propto t^{1/2}$ whatever the parameters are, so the measured exponent tests only that the system is on the attractor at all. And since $\Amp$ appears in Eq.~\eqref{eq:attractor_preview} only multiplied by $\ld$, the amplitude can be extracted from a density history only if $\ld$ is known independently. The second point is the reason for using this particular dataset. Direct annihilation and ordinary loop chopping both remove length at a rate $\propto\vbar/L$, so a one-scale average cannot separate them, and its single coefficient is a sum, $\tc\equiv\tceff=\tc_{\rm loop}+\tc_{\rm ann}$. What can be measured is $\Amp$, and only given $\ld$\footnote{The quantity we call $\ld$ is written $\Gamma$ in Eqs.~(6.14)--(6.24) of Ref.~\cite{Martins2016} and $\ld$ in its Eq.~(6.38); we reserve $\Gamma$ for the drag per unit length per unit velocity, following Ref.~\cite{Chuang1991PRL}, so that $\ld=T/\Gamma$ with $T$ the line tension.}. Reference~\cite{Chuang1991PRL} supplies $\ld$ on the same recordings, from the collapse of isolated, nearly circular defect loops, which obey $r^{2}=2\ld(t_{0}-t)$ for a loop of radius $r$ vanishing at time $t_{0}$; that paper verifies the predicted exponent at $0.50\pm0.03$. Because the same $\ld$ governs both the loop collapse and the network coarsening, the ratio of the two is free of the material constants of 5CB. This is why \CTY\ remains the relevant dataset despite far better modern imaging~\cite{ZushiTakeuchi2022}: it is the only source we know of that reports network coarsening and isolated-loop collapse for the same sample.
\begin{figure}[t]
  \centering
  \begin{tikzpicture}[>=Stealth,scale=0.86]
    \begin{scope}
      \foreach \s in {0,1,2}{
        \draw[very thick,domain=0:3.35,samples=120,smooth,variable=\x]
          plot ({\x},{\s+0.28*sin(112.5*\x)});}
      \draw[gray!65,dashed] (0.8,0.35) circle (0.93);
      \draw[<->,gray!65] (0.8,0.35) -- (0.8,1.28) node[midway,left=2pt]{\footnotesize $R$};
      \draw[<->] (3.2,1.0) -- (3.2,2.0) node[midway,right=1pt]{\footnotesize $L$};
      \node at (1.6,-1.35){\footnotesize (a) $\lam=R/L\simeq1$};
    \end{scope}
    \begin{scope}[shift={(4.55,0)}]
      \foreach \s in {0,1,2}{
        \draw[very thick,domain=0:3.35,samples=220,smooth,variable=\x]
          plot ({\x},{\s+0.197*sin(225*\x)});}
      \draw[gray!65,dashed] (0.4,0.867) circle (0.33);
      \draw[<->,gray!65] (0.4,0.867) -- (0.4,1.197) node[midway,left=2pt]{\footnotesize $R$};
      \draw[<->] (1.6,1.0) -- (1.6,2.0) node[midway,right=1pt]{\footnotesize $L$};
      \node at (1.6,-1.35){\footnotesize (b) $\lam\simeq1/3$};
    \end{scope}
  \end{tikzpicture}
  \caption{What the curvature ratio $\lam=R/L$ means, and why it is the whole uncertainty of this paper. Both panels show a tangle with the \emph{same} mean spacing $L$ between lines, marked by the black arrow. They differ only in how tightly the lines curve, measured by the radius $R$ of the dashed circle that best matches a line at its crest. (a) $\lam\simeq1$: lines curve on the scale of their own separation, the case $R=L$ usually assumed. (b) $\lam\simeq1/3$: the same spacing, but lines that bend on a scale finer than the distance to their neighbours. Since $\Amp=k(k+\tc\lam)/\lam^{2}$, the second tangle produces a far larger amplitude at the same $\tc$. This is why the measured amplitude of Sec.~\ref{sec:results} maps onto a curve in the $(\lam,\tc)$ plane rather than a value of $\tc$, and why deciding between panels of this kind requires imaging the tangle rather than its density.}
  \label{fig:lambdadef}
\end{figure}
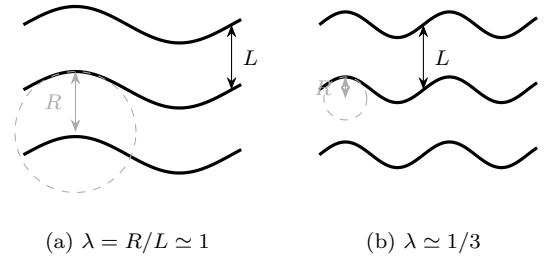

The remainder of this paper is organized as follows. In Sec.~\ref{sec:model} we evaluate the VOS model in the overdamped limit, isolate the amplitude as the sole measurable quantity, and derive the resulting bound on $\tc$. In Sec.~\ref{sec:fit} we state the assumptions connecting the theoretical attractor to the \CTY\ data, build the likelihood that handles the uncertainty in $\ld$, and justify grouping the runs by cell depth. In Sec.~\ref{sec:results} we present the fits. Section~\ref{sec:disc} discusses the systematics, compares against relativistic $\Uone$ calibrations, examines in detail why the two groups disagree, and lists what a repeat experiment would need to measure. Appendix~\ref{app:data} details the digitization and the point-selection rule; Appendix~\ref{app:matched} presents a test of whether the difference between the two groups can be attributed to any measured property of the runs.

\section{What is measurable}\label{sec:model}

\subsection{The symbols and what they mean} \label{sec:glossary}

The VOS model involves five quantities, listed in Table~\ref{tab:notation} and described here in turn. $L$ is the characteristic scale of the tangle, the mean spacing between neighbouring lines, defined through the length of line per unit volume as $1/L^{2}$. It is not what the experiment records\cite{Chuang1991PRL}: the published $\rho$ counts lines crossing unit area of an imaging plane, and converting a planar count into a length per unit volume is a stereological problem, treated as assumption~(iv) of Sec.~\ref{sec:assumptions}. $L$ is often called the correlation length, and we use the two terms interchangeably. $R$ is the typical radius of curvature of a line: large $R$ means gently curved lines, small $R$ means tightly kinked ones. The ratio $\lam\equiv R/L$ compares the two. At $\lam=1$ the lines curve on the scale of their own separation; at $\lam<1$ they bend on a scale finer than the distance to their neighbours. Two tangles with the same $L$ but different $\lam$ are shown in Fig.~\ref{fig:lambdadef}. The data analysed here do not determine $\lam$, and it is the dominant uncertainty in the result. $k$, the momentum parameter, measures how efficiently curvature is converted into motion. It is a correlation between the local velocity of a segment and the direction in which its curvature is pulling it, so a segment moving exactly along that direction, at uniform speed, has $k=1$. $\tc$, the sink coefficient, sets how fast the network loses length when segments meet. It is the coefficient of the second term of Eq.~\eqref{eq:VOS}, and is a pure number: the length lost per encounter divided by $L$. $\ld$ is the transport coefficient of the medium, with dimensions of area per unit time. It sets the overall pace at which anything happens, without influencing the shape of the time dependence. Of these, $L$ and $\ld$ are measured. The remaining three are what the VOS model would like to determine, and the burden of this section is that a density history can determine only one combination of them.

\begin{table}[t]
  \centering
  \caption{Notation, with the role each quantity plays in the present analysis. Only $\Amp$ is measured: $k$, $\lam$ and $\tc$ are never determined
  individually here.}
  \label{tab:notation}
  \begin{ruledtabular}
  \begin{tabular}{ll}
   Symbol & Meaning (and role) \\
   \hline
   $L$                     & mean spacing of lines (measured) \\
   $R$                     & typical curvature radius (not measured) \\
   $\lam\equiv R/L$        & tangle shape; wiggliness (free) \\
   $k$                     & curvature-to-motion efficiency ($k\le1$) \\
   $\kap\equiv k/\lam$     & the combination that appears \\
   $\tc$                   & length lost per encounter, in units of $L$ \\
   $\ld\equiv T/\Gamma$    & transport coefficient (area/time) \\
   $\Amp=\kap(\kap+\tc)$   & attractor amplitude (measured) \\
   $\vbar$                 & RMS velocity (eliminated) \\
   $\rho$                  & lines crossing unit area (the data) \\
   $\nu$                   & fitted exponent of $\rho(t)$; predicted $1$ \\
   $f_{j}$                 & reading error of one digitized point \\
   $s_{\ell}$              & prior width on the assigned $\ld$ \\
  \end{tabular}
  \end{ruledtabular}
\end{table}

\subsection{The overdamped attractor} \label{sec:attractor}

The VOS equations in flat space at fixed temperature are~\cite{MartinsShellard1996,Martins2016} 
\begin{equation}
  2\frac{dL}{dt}=\frac{L\vbar^{2}}{\ld}+\tc\,\vbar,
  \qquad
  \frac{1}{c^{2}}\frac{d\vbar}{dt}
   =\left(1-\frac{\vbar^{2}}{c^{2}}\right)\!
    \left[\frac{k}{R}-\frac{\vbar}{\ld}\right],
  \label{eq:VOS}
\end{equation}
with $c$ the speed of light. The two terms on the right of the first equation are exactly the two channels of Sec.~\ref{sec:picture}: the first is curve shortening, the second is removal at contact. The second equation is a force balance, with the curvature pull $k/R$ opposed by the drag term $\vbar/\ld$.

The one-scale hypothesis does not require $R=L$. It assumes only that every length in the network is proportional to a single dynamical scale, with time-independent ratios\footnote{Reference~\cite{Martins2016} implements it as $R\equiv L$, below its Eq.~(6.17); Ref.~\cite{Chuang1991PRL}, postulating the same single scale, states only that the curvature radius and the interstring separation are both \emph{proportional} to it.}. We therefore write $R=\lam L$ and keep $\lam$ explicit. Since $L$ is defined by the density and never by $R$, the ratio $\lam$ enters Eq.~\eqref{eq:VOS} only through the curvature term, where it is exactly degenerate with $k$: the combination $k/R$ equals $\kap/L$ with $\kap\equiv k/\lam$, so no measurement sensitive only to $k/R$ can tell the two apart\footnote{The equations are form-invariant and the one-scale description survives $\lam\ne1$ intact; what does not survive is the separate identification of $k$ and $\lam$. Note also that Eq.~\eqref{eq:VOS} is scale-free: $\ld$ is a diffusivity and no length appears in the equations, so on the attractor $\lam$ is a property of the material and the dynamical regime, and should be common to different quenches of the same substance unless the apparatus supplies a length of its own. In the data used here it does---the four runs were taken in cells of two different depths---and we return to this in Sec.~\ref{sec:fit}. By contrast $k$ and $\tc$ are properties of the line dynamics and of the topology of the vacuum manifold, and would ordinarily be common to all quenches of the same material regardless of the apparatus.}.

Disclination inertia lies far below the drag~\cite{ImuraOkano}, so the left-hand side of the velocity equation vanishes: the lines are always at terminal velocity. Setting the bracket in Eq.~\eqref{eq:VOS} to zero gives $\vbar=\kap\ld/L$, and substituting this into the first equation of \eqref{eq:VOS} gives
\begin{equation}
  \frac{d(L^{2})}{dt}=\Amp\,\ld,
  \qquad
  \Amp\equiv\kap(\kap+\tc)=\frac{k(k+\tc\lam)}{\lam^{2}}.
  \label{eq:attractor}
\end{equation}
The right-hand side is a constant, so $L^{2}$ grows linearly in time,
\begin{equation}
  L^{2}(t)=\Amp\,\ld\,t+L_{0}^{2},
  \label{eq:Lsq}
\end{equation}
with $L_{0}$ the scale at formation. The initial condition decays as $1/t$ relative to the growing term, which is the sense in which Eq.~\eqref{eq:attractor} describes an attractor: the memory of $L_{0}$ is erased. The two contributions to $\Amp$ are the two channels of Sec.~\ref{sec:picture}: curve shortening supplies $\kap^{2}$ and removal at contact supplies $\kap\tc$, so the amplitude measures how much faster the tangle coarsens than it would with no sink at all. At $\kap=1$ that factor is $1+\tc$.

Two consequences follow. First, the exponent implied by Eq.~\eqref{eq:Lsq} is $1/2$ for every value of $k$, $\tc$ and $\lam$, since all three sit inside the constant $\Amp$ and none appears in the power of $t$. The measured $\rho\propto t^{-1.02\pm0.09}$~\cite{Chuang1991PRL} therefore tests only that the system is on the overdamped attractor. Second, the ratio of the two channels is the constant $\tc/\kap$, so a likelihood built on $\rho(t)$ depends on $(k,\lam,\tc)$ only through $\Amp$. Its surfaces of constant $\chi^{2}$ in that three-dimensional space do not close around a best-fit point: an infinite family of parameter triples fits any dataset equally well. Inverting
Eq.~\eqref{eq:attractor} gives that family,
\begin{equation}
  \tc(\lam)=\frac{\Amp\lam}{k}-\frac{k}{\lam},
  \label{eq:ctilde}
\end{equation}
and reporting this curve rather than a number for $\tc$ is the strongest statement the data support.

\subsection{Bounding the momentum parameter} \label{sec:kbound}

The curve of Eq.~\eqref{eq:ctilde} can be converted into a one-sided bound on $\tc$, because $k$ is bounded above. From its definition (Eq.~(6.18) of Ref.~\cite{Martins2016}),
\begin{equation}
  \langle(1-\dot{\bm x}^{2}/c^{2})(\dot{\bm x}\cdot\hat{\bm u})\rangle
  \equiv k\,\vbar\,(1-\vbar^{2}/c^{2}),
\end{equation}
where $\bm x(s,t)$ is the position along the line, $s$ is arc length, and $\hat{\bm u}/R=d^{2}\bm x/ds^{2}$ is the unit vector pointing toward the centre of curvature. Since $|\hat{\bm u}|=1$ and $\langle|\dot{\bm x}|\rangle\le\vbar$, the non-relativistic limit obeys $k\le1$, with equality only if every part of the line moves exactly along $\hat{\bm u}$ at uniform speed. In words: $k$ cannot exceed unity because the velocity component along the curvature direction cannot exceed the total speed.

Both conditions hold on an isolated circular loop, so $k_{\rm loop}=1$ exactly, and the collapse of an isolated loop obeys $r^{2}=2k_{\rm loop}\ld(t_{0}-t)$. For the nearly circular loops measured in Ref.~\cite{Chuang1991PRL}, taking $k_{\rm loop}\simeq1$ is a reasonable approximation. For a tangled network no equivalent bound is available. The value $\knr=2\sqrt2/\pi\simeq0.900$ quoted in Eq.~(6.37) of Ref.~\cite{Martins2016} is the zero-velocity limit of a fitting form calibrated on relativistic Nambu--Goto simulations, and is therefore not a friction-regime calibration; the same reference suggests $k\simeq1$ in the condensed-matter regime. We therefore keep $k$ explicit throughout, quoting every number and curve at $k=1$ and reporting the value at $k=\knr$ alongside, so that the effect of lowering $k$ can be seen directly.

Because Eq.~\eqref{eq:ctilde} decreases monotonically with $k$, setting $k=1$ gives the smallest $\tc$ compatible with the data at each $\lam$:
\begin{equation}
  \tc\ \ge\ \Amp\lam-\frac{1}{\lam},
  \label{eq:floor}
\end{equation}
which at $\lam=1$ reads $\tc\ge\Amp-1$. The bound is vacuous for $\lam\le\Amp^{-1/2}$ and weakens continuously below $\lam=1$; in plain terms, below that value of $\lam$ the data say nothing about $\tc$ at all. This is the only place where the strict one-scale assumption does real work, and it should be noted that Eq.~\eqref{eq:floor} follows from $k\le1$, which converts to $\kap\le1$ only when $\lam\ge1$.

Finally, the amplitude has a direct reading as an energy-loss budget. On the attractor the fraction of $d(L^{2})/dt$ supplied by the sink rather than by curve shortening is $\tc\kap/\Amp=1-k^{2}/(\Amp\lam^{2})$. So at fixed $\kap$, a large $\Amp$ says that removal at contact supplies most of the coarsening rate. The qualification matters, since $\Amp$ can equally be raised at fixed $\tc$ by lowering $\lam$.

\section{Assumptions and likelihood}
\label{sec:fit}
Three limitations are intrinsic to what follows, and we state them  so that no result below is read as stronger than it is.

\begin{enumerate}[label=(\roman*),leftmargin=*]
\item The loop-collapse rate is published only as a range over a pressure interval~\cite{Chuang1991PRL}, and is not matched to individual quenches.
\item $\Amp$ constrains $k$, $\tc$ and $\lam$ only in the combination $\kap(\kap+\tc)$, so no analysis of the density can return $\tc$ by itself.
\item The four published quenches do not share one amplitude. We exclude none of them, reporting the three taken in the thicker cell as one group and the fourth
on its own.
\end{enumerate}

\subsection{The five assumptions}
\label{sec:assumptions}

Turning Eq.~\eqref{eq:attractor} into a prediction for the densities of Fig.~2 of Ref.~\cite{Chuang1991PRL} requires five assumptions\footnote{Reference \cite{ChuangPRE1993} complements Ref.~\cite{Chuang1991PRL} by extending the analysis to other defects, including loops, point defects, and event statistics. In both studies the experiments began from an initially isotropic state at a pressure of $3.6$~MPa, followed by quenches of varying magnitude $\Delta P$ across the isotropic-to-nematic phase transition. The reported initial temperature differs: $\sim33~^\circ$C in Ref.~\cite{Chuang1991PRL} versus $\sim37~^\circ$C in Ref.~\cite{ChuangPRE1993}.}. We list them separately so that each can be assessed on its own.

\begin{figure}[t]
  \centering
  \begin{minipage}{0.85\columnwidth}
    \centering
    \includegraphics[width=\linewidth]{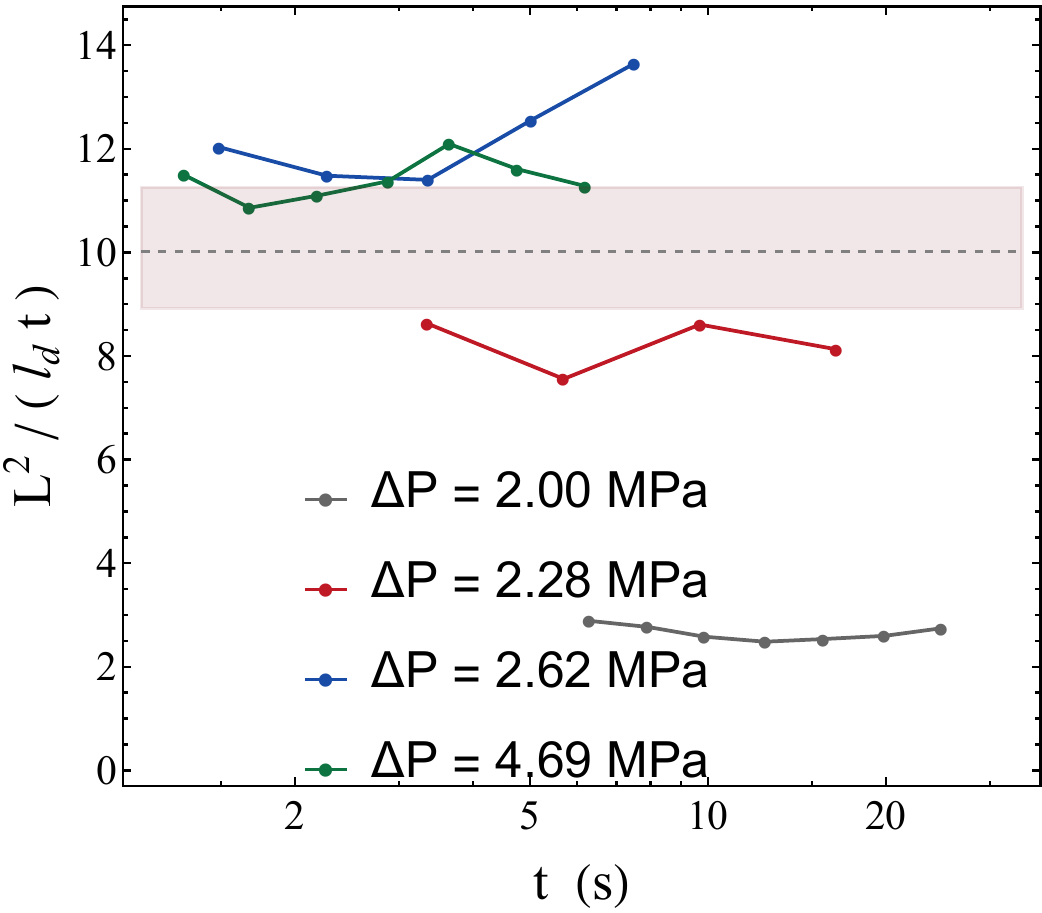}\\[0.3ex]
    (a)
  \end{minipage}\\[1.5ex]
  \begin{minipage}{0.85\columnwidth}
    \centering
    \includegraphics[width=\linewidth]{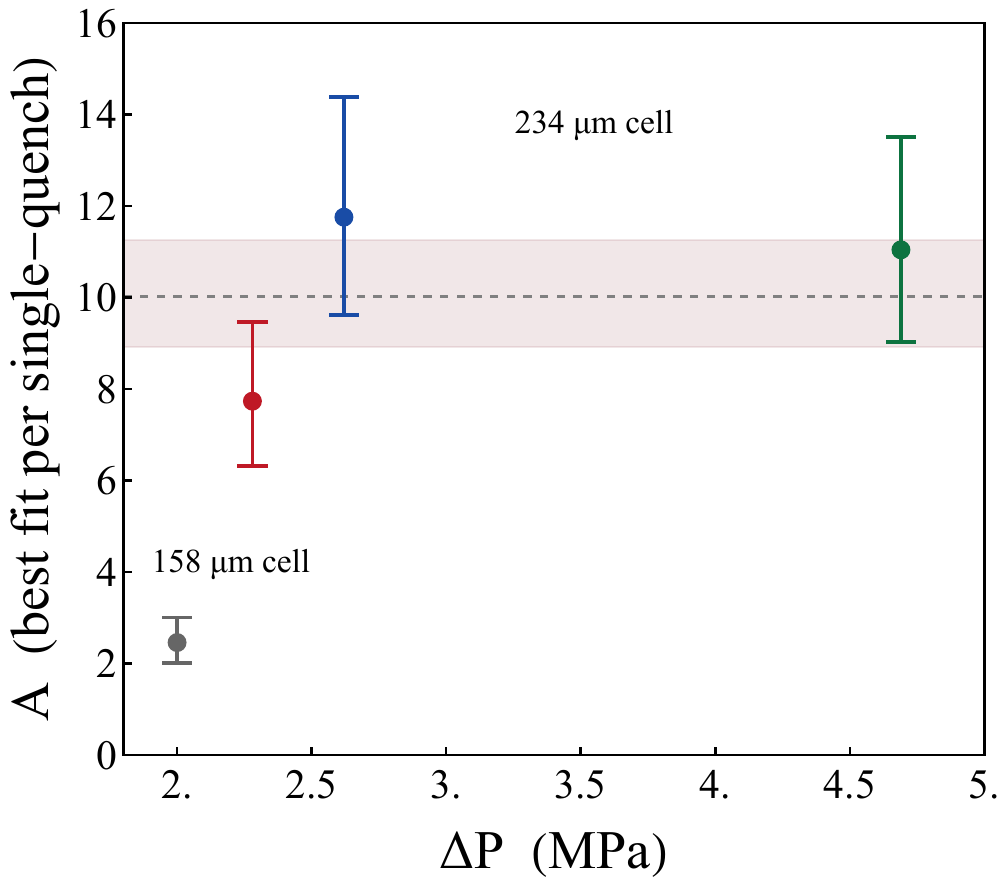}\\[0.3ex]
    (b)
  \end{minipage}
   \caption{(a) The point-by-point observable $\Aobs=L^{2}/(\ld t)$, obtained by applying Eq.~\eqref{eq:Lsq} to each datum separately with $L_{0}$ dropped. It is constant in time if the network is on the attractor, so this panel is a direct test of Eq.~\eqref{eq:Lsq} that involves no fitting. The three thick-cell runs give $7.6$--$8.6$, $11.4$--$13.7$ and $10.9$--$12.1$; the $\Delta P=2.00$~MPa run gives $2.5$--$2.9$ at a different amplitude. Constancy along each run is what the one-scale assumption asserts; equality between runs is what it does not. The absence of any systematic downward drift is consistent with assumption (iii), since a neglected $L_{0}$ would make $\Aobs$ fall monotonically with time; the residual scatter within each run is not monotonic and so does not have that form. (b) The single-quench amplitudes of Table~\ref{tab:single} against quench depth, with the prior-limited errors, labelled by cell thickness. Band and dashed line are the three-run result, Eq.~\eqref{eq:Ares}. Together the two panels are why the three thick-cell quenches are fitted jointly and the fourth on its own.}
  \label{fig:runs}
\end{figure}

\begin{enumerate}[label=(\roman*),leftmargin=*]
\item \textbf{Overdamped limit.} $d\vbar/dt=0$: the lines are always at terminal velocity, justified for disclinations by Ref.~\cite{ImuraOkano}.
\item \textbf{One-scale geometry.} $R=\lam L$ with $\lam$ constant within a run,
but otherwise free.
\item \textbf{Negligible initial condition.} $L_{0}$ is dropped from Eq.~\eqref{eq:Lsq}. Retaining it would give an apparent amplitude $\Aobs\equiv L^{2}/(\ld t)=\Amp+L_{0}^{2}/(\ld t)$, the value of $\Amp$ implied by one datum on its own, which decreases with time. No such drift is seen among the retained points of any run in Fig.~\ref{fig:runs}(a). The dropped first point of each run does lie high, but the early-time undercounting identified in Ref.~\cite{Chuang1991PRL} biases $\Aobs$ in the same direction, so the two cannot be separated by a density history; both are removed by the selection of Appendix~\ref{app:data}, which also gives the effect of retaining them.
\item \textbf{Isotropic conversion.} The published observable $\rho$ is the number of lines crossing a unit area of an imaging plane, as calibrated in Ref.~\cite{Chuang1991PRL}. Converting it to the length of line per unit volume, $\Lambda=1/L^{2}$, requires knowing how the lines are oriented relative to that plane; for an isotropic tangle of lines the standard stereological result is a factor of $2$, giving $\Lambda=2\rho$ \cite{Underwood1970}. 
\item \textbf{Transport coefficient transfers.} The $\ld$ governing the network is taken to be the one measured on isolated loops\footnote{A circular loop obeys $\Gamma\dot r=-T/r$, the $k_{\rm loop}=1$ case of $\vbar=\ld/R$, integrating to $r^{2}=2\ld(t_{0}-t)$, Eq.~(2) of Ref.~\cite{Chuang1991PRL}. That paper measures the collapse exponent as $0.50\pm0.03$ from seven events, and states that from its Eq.~(2) the quantity $T/\Gamma$ varied from about $200$ to $300~\mu$m$^{2}$/s in the $5.6$--$8.3$~MPa regime. The published range is therefore $\ld$ itself and not the slope $2\ld$; reading the endpoints of its Fig.~4, $r\simeq48~\mu$m at $t_{0}-t\simeq5$~s, gives $r^{2}/(t_{0}-t)\simeq4.6\times10^{2}~\mu$m$^{2}$/s and hence $\ld\simeq2.3\times10^{2}~\mu$m$^{2}$/s, consistent with that reading.}.
\end{enumerate}

Assumption (v) is what makes the whole measurement possible, and at the level of material constants it is exact. The line tension $T$ and the drag coefficient $\Gamma$ both diverge logarithmically with the ratio of the sample size to the defect core radius, and they do so identically, so their ratio $\ld=T/\Gamma=K/\gamma$ is independent of that cutoff. Here $K$ is the Frank elastic constant and $\gamma$ the rotational viscosity of the liquid crystal. The published range $200$--$300~\mu$m$^{2}$/s~\cite{Chuang1991PRL} refers to $\ld$ itself.

With $\rho$ in mm$^{-2}$, $t$ in s and $\ld$ in $\mu$m$^{2}$/s, combining assumptions (iii) and (iv) with Eq.~\eqref{eq:Lsq} gives the prediction
\begin{equation}
  \rho_{\rm th}(t)=\frac{10^{6}}{2\,\Amp\,\ld\,t_{*}}
                   \left(\frac{t}{t_{*}}\right)^{-\nu},
  \qquad t_{*}\equiv1~\text{s},
  \label{eq:rho}
\end{equation}
where the factor $10^{6}$ converts $\mu$m$^{-2}$ to mm$^{-2}$, and $t_{*}$ is a reference time, fixed at one second, at which the amplitude is defined. The exponent $\nu$ is predicted to be $1$ but is left free as a consistency check. 

\subsection{The data and the assigned transport coefficient} \label{sec:datasel}

We use the four quenches of Ref.~\cite{Chuang1991PRL} ($\Delta P=2.00$, $2.28$, $2.62$ and $4.69$~MPa) from an initial isotropic state at $P_{i}=3.6$~MPa, taken in sample cells of depth $d=158\pm8~\mu$m and $234\pm23~\mu$m. The pressure is raised, so the final pressure is $P_{f}=P_{i}+\Delta P$, giving $P_{f}=5.60$, $5.88$, $6.22$ and $8.29$~MPa. Throughout we call the single thin-cell run group~1 and the three thick-cell runs group~2. Following that paper we drop the first and last points of each run, and additionally the second point of the $\Delta P=2.00$~MPa quench, which begins later than the others; the rule and its effect are set out in Appendix~\ref{app:data}.

The loop-collapse measurement is reported only as a range across a span of final pressures, and is not matched to individual runs. We therefore assign each quench a value by interpolating linearly between the endpoints of that range:
\begin{equation}
  \bar{\ell}_d(P_{f})=200+\tfrac{100}{2.7}\,(P_{f}-5.6)~\mu\text{m}^{2}/\text{s},
  \label{eq:interp}
\end{equation}
giving $\{200,210,223,300\}~\mu$m$^{2}$/s for the four runs. This interpolation is an assumption of ours and not a measurement, which is why the uncertainty it carries is propagated explicitly through the likelihood below rather than ignored.

The $\Delta P=2.00$~MPa run differs from the other three in two ways at once: it is the only one taken in the $158~\mu$m cell, and it is the shallowest quench. Either could matter. A shallower quench forms a sparser initial tangle, and while the initial \emph{density} is erased on the attractor, the tangle \emph{shape} $\lam$ is not a quantity the model derives, and may retain memory of formation. A difference in $\lam$ would move $\Amp$ without moving $\nu$, which is exactly what is observed. We record this as a possibility and not a preference, and note that it is exactly the ambiguity that running all quenches in a single cell would remove. Appendix~\ref{app:matched} shows that the difference does not act through any of the measured properties the two groups share, and behaves instead as a constant offset attached to the run; Sec.~\ref{sec:disc} works through the candidate explanations.

\subsection{Why the likelihood has the form it does}
\label{sec:like}

It is convenient to work throughout with the logarithmic residual of a single
digitized point,
\begin{equation}
  \mathcal{R}^{(j)}_{i}\;\equiv\;\ln\rho^{(j)}_{i}-\ln\rho^{(j)}_{\rm th}\bigl(t^{(j)}_{i}\bigr),
\end{equation}
the amount by which point $i$ of quench $j$ exceeds the prediction for that quench, measured in natural-log units so that it is a fractional rather than an absolute discrepancy. Everything below is a statement about the $\mathcal{R}^{(j)}_{i}$.

Two distinct errors enter them, and they act differently. The reading error $f_{j}$ of a single point scatters the residuals of a quench \emph{about their own mean}. The error on the assigned $\bar\ld_{j}$ of Eq.~\eqref{eq:interp} is common to the whole quench and \emph{displaces that mean}, up or down, without changing the slope. Treating the second as if it were the first would be wrong in both directions: it would overstate the evidence coming from any one quench, and understate the uncertainty on the amplitude.

A way to handle this is to give each quench its own offset $\delta_{j}$, drawn from a distribution of width fixed in advance---what is called a random-effects structure.  Its only purpose is to combine the three quenches of the $234~\mu$m cell into a single amplitude without counting them as three independent measurements of it. Each has its own uncertain $\bar\ld_{j}$, and uncertainties of that kind do not average down the way point-to-point scatter does; the second term of Eq.~\eqref{eq:chi2} is what stops them from doing so. The $\Delta P=2.00$~MPa quench is fitted alone, and for it none of this is needed. We take both errors to be log-normal, meaning Gaussian in the logarithm, so that they act multiplicatively on the density rather than additively, and write
\begin{equation*}
  \mathcal{R}^{(j)}_{i}=\delta_{j}+\varepsilon^{(j)}_{i},\qquad
  \varepsilon^{(j)}_{i}\sim N\bigl(0,f_{j}^{2}\bigr),\qquad
  \delta_{j}\sim N\bigl(0,s_{\ell}^{2}\bigr),
\end{equation*}
independently over points $i$ and over quenches $j$, where $N(m,v)$ denotes a Gaussian of mean $m$ and variance $v$. A distribution specified in advance in this way, before the density data are looked at, is called a prior; the width $s_{\ell}$ is the prior on the per-quench offset, and encodes how far we believe the assigned $\bar\ld_{j}$ may be from the truth.

From here we work with $\chi^{2}\equiv-2\ln\mathcal{L}$, twice the negative logarithm of the likelihood of the model above, so that minimizing $\chi^{2}$ maximizes the probability of the observed densities and intervals can be read off from the rise of $\chi^{2}$ above its minimum. For one quench this gives

\begin{equation}
  \chi^{2}_{(j)}(\delta_{j})
  =\frac{1}{f_{j}^{2}}\sum_{i}\bigl(\mathcal{R}^{(j)}_{i}-\delta_{j}\bigr)^{2}
  +\frac{\delta_{j}^{2}}{s_{\ell}^{2}},
\end{equation}
in which the two terms have a plain reading: the first asks the points to agree with the model once the whole quench has been shifted by $\delta_{j}$, and the second is the price of that shift. Because the expression is a strictly convex quadratic, it has a single minimum, at
\begin{equation}
  \hat\delta_{j}=\frac{s_{\ell}^{2}S^{(j)}}{f_{j}^{2}+n_{j}s_{\ell}^{2}}
  =\frac{n_{j}s_{\ell}^{2}}{f_{j}^{2}+n_{j}s_{\ell}^{2}}\,\mathcal{\bar R}^{(j)},
  \qquad S^{(j)}\equiv\sum_{i}\mathcal{R}^{(j)}_{i},
\end{equation}
where $\mathcal{\bar R} ^{(j)}=S^{(j)}/n_{j}$ is the mean residual of the quench and $n_{j}$ its number of points. The prefactor lies between zero and one, so the fitted offset is the mean residual \emph{shrunk} toward zero, which is where the prior expects it to be; the word is literal, and the amount of shrinkage is set by how far the prior trusts the assigned $\ld$.

We do not want the $\delta_{j}$ in the final answer, so we integrate them out---marginalization, which propagates their uncertainty into the error on $\Amp$ rather than leaving it hidden. Substituting the shrunk offset and summing over quenches,
\begin{equation}
  \chi^{2}(\Amp,\nu)=\sum_{j}\left[
  \frac{1}{f_{j}^{2}}\sum_{i}\bigl(\mathcal{R}^{(j)}_{i}\bigr)^{2}
  -\frac{s_{\ell}^{2}\bigl(S^{(j)}\bigr)^{2}}{f_{j}^{2}\,\bigl(f_{j}^{2}+n_{j}s_{\ell}^{2}\bigr)}\right].
  \label{eq:chi2}
\end{equation}
The first term is the ordinary diagonal $\chi^{2}$. The second removes the part of the mismatch in which every point of a quench deviates in the same direction---exactly the pattern a wrong $\bar\ld_{j}$ produces---but removes it only partially, in proportion to $s_{\ell}^{2}$, so that a common offset much larger than the published $20\%$ still costs. The two limits confirm the structure: at $s_{\ell}=0$ the second term vanishes and Eq.~\eqref{eq:chi2} reverts to the diagonal $\chi^{2}$, while as $s_{\ell}\to\infty$ it becomes $\bigl(S^{(j)}\bigr)^{2}/(n_{j}f_{j}^{2})$, leaving $\sum_{i}\bigl(\mathcal{R}^{(j)}_{i}- \mathcal{\bar R}^{(j)}\bigr)^{2}/f_{j}^{2}$---the normalization of the quench is then free, and only the time dependence of $\rho$ carries information. Because the Gaussian integral over each $\delta_{j}$ contributes a factor independent of $\Amp$, $\nu$ and the data, Eq.~\eqref{eq:chi2} is simultaneously the profile likelihood (maximized over the offsets) and the marginal likelihood (integrated over them), so nothing turns on which convention the reader prefers.

For the reading errors $f_{j}$ we take the plotted symbol half-size in natural-log units: $0.056$ for the three thick-cell quenches and $0.050$ for the plus symbols of the $\Delta P=2.00$~MPa quench. Reference~\cite{Chuang1991PRL} states only that its statistical errors are smaller than the symbols, so treating that upper limit as a Gaussian $1\sigma$ is a convention, which also absorbs the error made in digitizing. Since the error on the exponent is set by the $f_{j}$ alone, the $\sigma_{\nu}$ quoted below should be read as a lower limit. The amplitude, by contrast, is controlled by the prior rather than by the scatter of the points: it moves by under $2\%$ if the $f_{j}$ are doubled.

For the prior width we take $s_\ell=0.2\simeq\frac12\ln(300/200)$, that is, we read the published range as a $\pm1\sigma$ interval about its geometric centre. The resulting error on $\Amp$ is therefore essentially a restatement of that range. 
\subsection{Properties of the likelihood} \label{sec:props}

\emph{The error interval is exact, not asymptotic.} Because $\rho_{\rm th}\propto1/\Amp$ in Eq.~\eqref{eq:rho}, the amplitude enters the prediction as a uniform multiplicative factor, so in the logarithm it shifts every residual of every run by the same amount. Each residual therefore splits as
\begin{equation}
  \mathcal{R}^{(j)}_{i}=a^{(j)}_{i}+\ln\Amp,
  \quad
  a^{(j)}_{i}\equiv\ln\rho^{(j)}_{i}
  -\ln\!\left[\frac{10^{6}}{2\,\bar\ld_{j}t_{*}}\right]
  +\nu\ln\!\left(\frac{t^{(j)}_{i}}{t_{*}}\right),
  \label{eq:residsplit}
\end{equation}
where $a^{(j)}_{i}$ collects the measured density, the assigned transport coefficient and the time dependence, and is independent of $\Amp$. Hence $\partial\mathcal{R}^{(j)}_{i}/\partial\ln\Amp=1$ at every point of every run. Introducing the two data sums of run $j$, $P_{j}\equiv\sum_{i}a^{(j)}_{i}$ and $Q_{j}\equiv\sum_{i}\bigl(a^{(j)}_{i}\bigr)^{2}$, both independent of $\Amp$, the two ingredients of Eq.~\eqref{eq:chi2} become
$\sum_{i}(\mathcal{R}^{(j)}_{i})^{2}=Q_{j}+2P_{j}\ln\Amp+n_{j}\ln^{2}\!\Amp$ and $S^{(j)}=P_{j}+n_{j}\ln\Amp$, both of degree exactly two in $\ln\Amp$. Consequently $\chi^{2}=G\ln^{2}\!\Amp+B\ln\Amp+C$, where $G$, $B$ and $C$ denote the coefficients of the quadratic, linear and constant terms; of these only $G$ is needed, since $B$ merely fixes the position of the minimum and $C$ cancels in $\Delta\chi^{2}$. Collecting the terms in $\ln^{2}\!\Amp$,
\begin{equation}
  G=\sum_{j}\frac{n_{j}}{f_{j}^{2}}
  \left[1-\frac{n_{j}s_{\ell}^{2}}{f_{j}^{2}+n_{j}s_{\ell}^{2}}\right]
  =\sum_{j}\frac{n_{j}}{f_{j}^{2}+n_{j}s_{\ell}^{2}}.
  \label{eq:curvature}
\end{equation}
Neither $P_{j}$ nor $Q_{j}$ survives: they enter only $B$ and $C$, so the curvature is fixed by the number of retained points and not by their values. Completing the square then gives $\chi^{2}-\chi^{2}_{\min}=G\,(\ln\Amp-\ln\hat\Amp)^{2}$, with $\ln\hat\Amp=-B/(2G)$ the best fit, an identity rather than a second-order expansion about the minimum, so $\Delta\chi^{2}=1$ gives $|\ln\Amp-\ln\hat\Amp|=G^{-1/2}$ exactly and hence
\begin{equation}
  \sigma_{\ln\Amp}=\left[\sum_{j}\frac{n_{j}}{f_{j}^{2}+n_{j}s_{\ell}^{2}}\right]^{-1/2},
  \label{eq:sigma}
\end{equation}
which is exact and symmetric in $\ln\Amp$, being the usual $68\%$ interval for a single parameter, rather than an approximation valid only for large samples. That distinction is not academic here: with four runs and twenty-three retained points, an interval justified only in the large-sample limit would be doing real work.

Because the interval is symmetric in the logarithm of the amplitude and not in the amplitude itself, it maps to an asymmetric interval on $\Amp$, which is the origin of the unequal error bars quoted throughout. The limits are simply $\Amp\,e^{\pm\sigma_{\ln\Amp}}$: for group~2 with $\sigma_{\ln\Amp}=0.116$ this turns $\Amp=10.46$ into $^{+1.29}_{-1.15}$, and the same rule generates every other interval in Table~\ref{tab:fits}. Releasing the exponent widens $\sigma_{\ln\Amp}$ by $(1-{\rm corr}^{2})^{-1/2}$, where ${\rm corr}$ is the correlation coefficient between amplitude and exponent.

\emph{A single quench measures the amplitude only up to its own offset.} Within one quench, $\Amp$ and $\delta_{j}$ shift $\ln\rho^{(j)}_{\rm th}$ by the same constant at every point, so the data cannot distinguish them. Setting the derivative of Eq.~\eqref{eq:chi2} with respect to $\ln\Amp$ to zero gives $2S^{(j)}/(f_{j}^{2}+n_{j}s_{\ell}^{2})=0$, that is $S^{(j)}=0$, in which $s_{\ell}$ does not appear: the best fit is the amplitude that makes theresiduals of that quench sum to zero, whatever the prior width may be. The prior sets only the error quoted beside it, which is then a restatement of the published $200$--$300~\mu$m$^{2}$/s range and not a measurement of scatter. That error is nonetheless the right one to quote: the alternative, the point-to-point error $f_{j}/\sqrt{n_{j}}$ obtained at $s_{\ell}=0$, runs from $1.9\%$ for the $\Delta P=2.00$~MPa quench to $2.8\%$ for the $\Delta P=2.28$~MPa one, roughly an order of magnitude below the $20\%$ uncertainty on the $\ld$ that multiplies the amplitude. This is why the amplitude of the $\Delta P=2.00$~MPa quench in Sec.~\ref{sec:results} carries an error of $\pm20\%$; only a second quench in the same cell could improve on it.

\emph{The information saturates: adding density points cannot improve $\Amp$.} Each quench contributes $n_{j}/(f_{j}^{2}+n_{j}s_{\ell}^{2})$ to the sum in Eq.~\eqref{eq:sigma}---a contribution to the inverse variance of $\ln\Amp$, so that a larger value means a tighter constraint---and this tends to $1/s_{\ell}^{2}$ as the number of points $n_{j}$ grows, however many are added. With $s_{\ell}=0.2$ the ceiling is therefore $25$ per quench, and the three thick-cell quenches already give $24.5$, $24.6$ and $24.7$. Hence $\sigma_{\ln\Amp}=0.116\simeq s_{\ell}/\sqrt3$ for that group and $0.201\simeq s_{\ell}$ for the single quench.  The limiting quantity is the number of \emph{quenches} with an independently known $\ld$, not the number of points on each.

\section{Results} \label{sec:results}

Before the numbers, one statement about their form. Because $\chi^{2}$ depends on $k$, $\lam$ and $\tc$ only through $\Amp$, the output of this analysis is a curve in parameter space and not a number for $\tc$. Every point on that curve fits the data equally well. The amplitudes below are measurements; the values of $\tc$ derived from them are conditional on a choice of $\lam$ that the data do not supply.

\begin{table}[t]
  \centering
  \caption{Single-quench fits, one run at a time, exponent free. Each entry is the amplitude implied by the assigned $\ld$ of Eq.~\eqref{eq:interp} taken at face value, and carries the prior width, $\pm20\%$, as its error. The three thick-cell quenches agree within a factor $1.3$; the thin-cell quench does not. The $\Delta P=2.62$~MPa exponent, $\nu=1.084$, is the furthest from unity of the four, but its offset from $1$ is under $1.5\sigma$ on the per-run error and it does not drive the joint fit.}
  \label{tab:single}
  \begin{ruledtabular}
  \begin{tabular}{lcccccc}
   $\Delta P$ & $P_{f}$ & $\ld$ & $d$ ($\mu$m) & $n$ & $\Amp$ & $\nu$ \\
   \hline
   $2.00$ & $5.60$ & $200$ & $158$ & $7$ & $2.98$  & $0.954$ \\
   $2.28$ & $5.88$ & $210$ & $234$ & $4$ & $8.36$  & $0.991$ \\
   $2.62$ & $6.22$ & $223$ & $234$ & $5$ & $11.02$ & $1.084$ \\
   $4.69$ & $8.29$ & $300$ & $234$ & $7$ & $11.14$ & $1.022$ \\
  \end{tabular}
  \end{ruledtabular}
\end{table}

\begin{table}[t]
  \centering
  \caption{Baseline fits, Eq.~\eqref{eq:chi2}, $s_{\ell}=0.2$. Errors correspond to a rise of one unit in $\chi^{2}$ and are exact via Eq.~\eqref{eq:sigma}. The two  groups are treated identically; the $\Delta P=2.00$~MPa error is the prior width because a single run cannot do better. $N$ is the number of points; the free parameters are $\Amp$ and, where indicated, $\nu$, the per-run offsets being integrated out rather than fitted, so the exponent-free rows correspond to $N-2$ effective degrees of freedom. Errors on the exponent-free rows are marginalized over $\nu$ and are wider than Eq.~\eqref{eq:sigma} by $(1-\mathrm{corr}^2)^{-1/2}$; the $\nu\equiv1$ rows use Eq.~\eqref{eq:sigma}  directly. The widening is $4.6\%$ for group~2 and $12.6\%$ for group~1, whose amplitude--exponent correlation is the stronger, $\mathrm{corr}=-0.459$ against $-0.293$. The derived rows use Eq.~\eqref{eq:ctilde} at $\lam=1$ and are quoted only to show the size of the $k$-degeneracy; they are not a measurement of
  $\tc$.}
  \label{tab:fits}
  \begin{ruledtabular}
  \begin{tabular}{lcccc}
   Fit & $\Amp$ & $\nu$ & $\chi^{2}_{\min}$ & $N$ \\
   \hline
   \multicolumn{5}{l}{Group 2: $\Delta P=2.28,2.62,4.69$~MPa ($234~\mu$m cell)} \\
   \ \ exponent free & $10.02^{+1.30}_{-1.15}$ & $1.030\pm0.025$ & $13.59$ & $16$ \\
   \ \ $\nu\equiv1$  & $10.46^{+1.29}_{-1.15}$ & $\equiv1$       & $15.03$ & $16$ \\
   \hline
   \multicolumn{5}{l}{Group 1: $\Delta P=2.00$~MPa ($158~\mu$m cell)} \\
   \ \ exponent free & $2.98^{+0.76}_{-0.60}$ & $0.954\pm0.041$ & $5.81$ & $7$ \\
   \ \ $\nu\equiv1$  & $2.65^{+0.59}_{-0.48}$ & $\equiv1$       & $7.08$ & $7$ \\
   \hline
   \multicolumn{5}{l}{Derived at $\lam=1$, exponent free:} \\
   \ \ $\tc$ at $k=1$    & \multicolumn{4}{l}{$9.02^{+1.30}_{-1.15}$ (gr.~2), $1.98^{+0.76}_{-0.60}$ (gr.~1)} \\
   \ \ $\tc$ at $k=\knr$ & \multicolumn{4}{l}{$10.23$ (gr.~2), $2.41$ (gr.~1)} \\
  \end{tabular}
  \end{ruledtabular}
\end{table}

\subsection{The two amplitudes} \label{sec:amps}

\begin{figure}[t]
  \centering
  \includegraphics[width=\columnwidth]{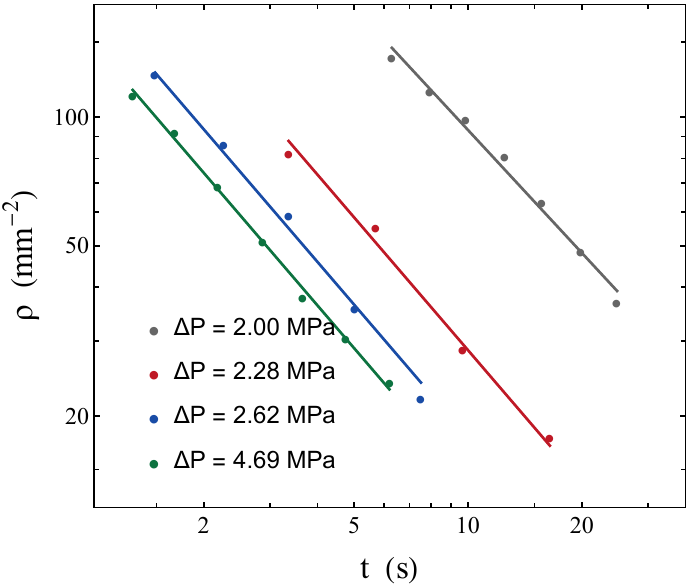}
  \caption{The four quenches, digitized from Fig.~2 of
  Ref.~\cite{Chuang1991PRL}. The three thick-cell curves share the single amplitude of Eq.~\eqref{eq:Ares} and the $\Delta P=2.00$~MPa curve that of  Eq.~\eqref{eq:Ares1}; within each group the curves differ only through the assigned $\ld$ of Eq.~\eqref{eq:interp} and the offsets $\hat\delta_{j}$, so they are not independent fits, and each is drawn only over the span of its own data. Error bars are omitted because Ref.~\cite{Chuang1991PRL} states only that its statistical errors are smaller than the plotted symbols; the reading errors $f_{j}$ are given in Table~\ref{tab:data}.}
  \label{fig:fit}
\end{figure}

The three quenches in the $234~\mu$m cell share an amplitude,
\begin{equation}
  \Amp_{2}=10.0^{+1.3}_{-1.1},
  \label{eq:Ares}
\end{equation}
with the exponent free and $\chi^{2}_{\min}=13.59$ for $16$ points and two free parameters. Imposing the predicted $\nu=1$ gives $10.5$ (Table~\ref{tab:fits}, Fig.~\ref{fig:fit}). The fitted per-run offsets are $\hat\delta_{j}=\{+0.25,-0.16,-0.10\}$, that is $\{+29,-15,-9\}\%$ in $\rho$, all lying within the prior, with the largest at $1.3\sigma$: the assigned $\ld$ values of Eq.~\eqref{eq:interp} are therefore not being strained by the fit.

The $\Delta P=2.00$~MPa quench, fitted on its own by the same likelihood, gives
\begin{equation}
  \Amp_{1}=3.0^{+0.8}_{-0.6},
  \label{eq:Ares1}
\end{equation}
where the error is the prior width and not a measurement of the scatter in the data.

\subsection{The exponent check} \label{sec:expcheck}

\begin{figure}[htbp]
  \centering
  \begin{minipage}{\columnwidth}
    \centering
    \includegraphics[width=\linewidth]{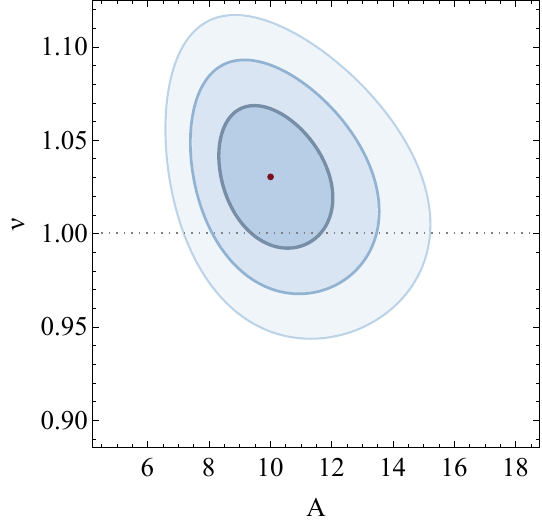}\\[0.3ex]
    (a)
  \end{minipage}\\[1.5ex]
  \begin{minipage}{\columnwidth}
    \centering
    \includegraphics[width=\linewidth]{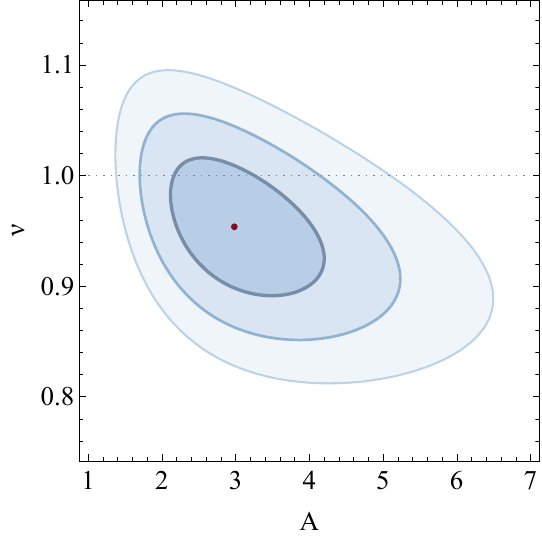}\\[0.3ex]
    (b)
  \end{minipage}
   \caption{Joint constraint on amplitude and exponent. Contours are at $\Delta\chi^{2}=2.30$, $6.18$ and $11.83$, the $68\%$, $95\%$ and $99.7\%$ levels for two parameters jointly. The dotted line marks the predicted $\nu=1$; the point is the best fit. (a) The three-run group, which gives $\nu=1.030\pm0.025$, recovering $\nu=1$ at $1.2\sigma$. The anticorrelation, ${\rm corr}=-0.29$, is fixed by the reference time $t_{*}=1$~s at which the amplitude is defined. (b) The $\Delta P=2.00$~MPa run, which gives $\nu=0.954\pm0.041$, recovered at $1.1\sigma$; its amplitude--exponent correlation is the stronger of the two, ${\rm corr}=-0.46$, which is why its ellipse is the more elongated. Each group is thus consistent with $\nu=1$, so the two differ in amplitude and not in exponent.}
  \label{fig:Anu}
\end{figure}

Both groups pass the same consistency check. Releasing $\nu$ returns $1.030\pm0.025$ for the three-run group and $0.954\pm0.041$ for the $\Delta P=2.00$~MPa run. The value expected for curvature-driven coarsening---the Lifshitz--Allen--Cahn law $L\propto t^{1/2}$, equivalently $\nu=1$ for the density---is thus recovered at $1.2\sigma$ and $1.1\sigma$ respectively, and imposing $\nu\equiv1$ costs $\Delta\chi^{2}=1.44$ and $1.27$. Figure~\ref{fig:Anu} shows the joint constraint on the two parameters. The amplitude and exponent are anticorrelated, ${\rm corr}(\Amp,\nu)=-0.29$ for the three-run group; the size of that correlation is not physical but is fixed by where the amplitude is defined, namely the reference time $t_{*}=1$~s in Eq.~\eqref{eq:rho}. Neither exponent improves on the published $1.02\pm0.09$. It is the same data with a different error model, and $\sigma_{\nu}$ is set by the $f_{j}$, which are symbol-size upper limits excluding the digitization error. Their role here is narrower: to show that Eqs.~\eqref{eq:Ares}--\eqref{eq:Ares1} are not artifacts of imposing $\nu=1$, and that the difference between the two groups is one of normalization and not of exponent. Both are on the overdamped attractor.

\subsection{The constraint curve}\label{sec:curve}

\begin{figure}[t]
  \centering
  \includegraphics[width=\columnwidth]{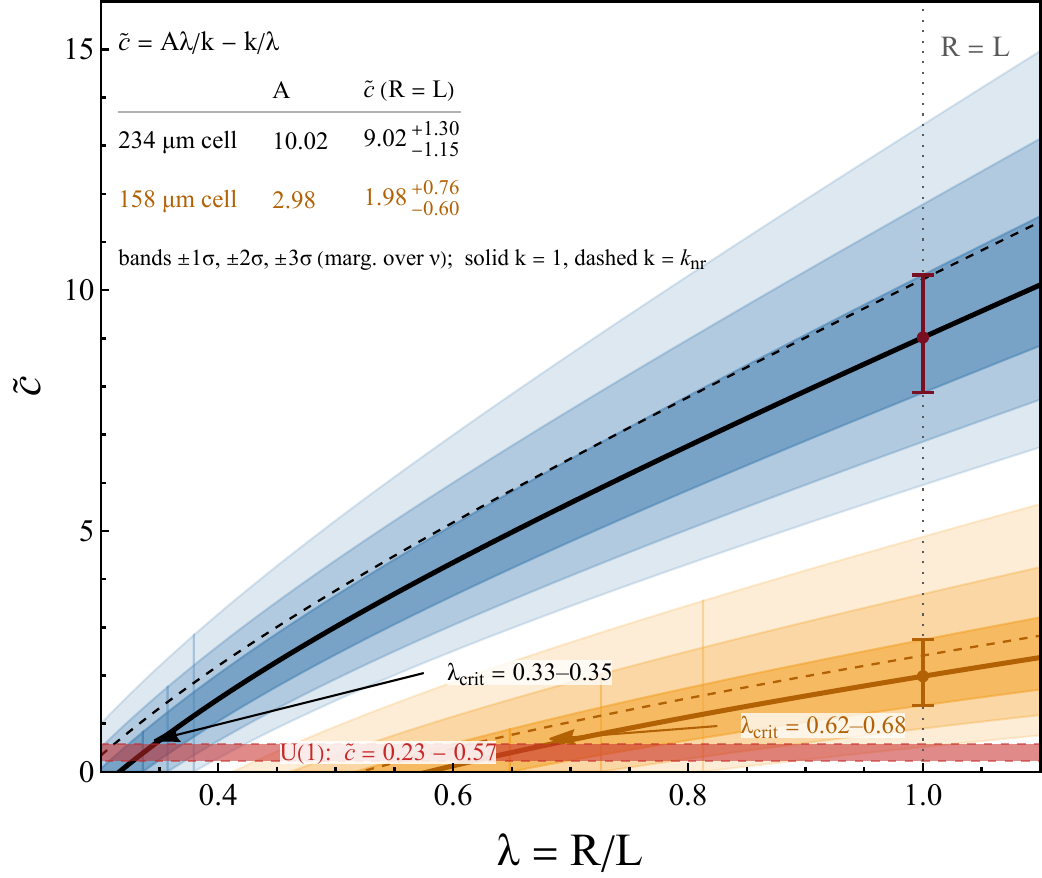}
  \caption{The result of this paper. \emph{Every point on a given curve fits the corresponding data equally well}: since $\chi^{2}$ depends on $k$, $\lam$ and $\tc$ only through $\Amp=k(k+\tc\lam)/\lam^{2}$, the data constrain them only along $\tc=\Amp\lam/k-k/\lam$. This is a one-parameter family of equally good fits and not a two-parameter measurement, so the bands should not be read as confidence contours in the $(\lam,\tc)$ plane. Black, the three quenches in the $234~\mu$m cell, $\Amp=10.02$; orange, the $\Delta P=2.00$~MPa quench in the $158~\mu$m cell, $\Amp=2.98$; the two are fitted separately and neither is preferred. Solid, $k=1$, which by $k\le1$ is the smallest $\tc$ compatible with the data at each $\lam$; dashed, $k=\knr=0.900$, spanning the whole momentum-parameter uncertainty. Bands are the $\Delta\chi^{2}=1$, $4$ and $9$ intervals on $\Amp$, marginalized over $\nu$, mapped through the same relation; they are intervals on one parameter and the figure closes in neither direction. The error bars at $\lam=1$ are $\tc(R=L)$ with its $1\sigma$ interval, $9.02^{+1.30}_{-1.15}$ and $1.98^{+0.76}_{-0.60}$. The strip is the oriented $\Uone$ calibration $\tc_{U(1)}=0.23$--$0.57$, reached at $\lam_{\rm crit}=0.33$--$0.35$ and $0.62$--$0.68$ respectively. The dotted vertical marks $R=L$.}
  \label{fig:lambda}
\end{figure}

Everything that follows comes from combining Eqs.~\eqref{eq:Ares} and~\eqref{eq:Ares1} with Eq.~\eqref{eq:ctilde}, and inherits the degeneracy of Sec.~\ref{sec:model}. Figure~\ref{fig:lambda} is the result of the paper. The bands are the errors on $\Amp$ mapped through Eq.~\eqref{eq:ctilde}.

At $\lam=1$ with $k\le1$ the curves give $\tc\ge9.0^{+1.3}_{-1.1}$ and $\tc\ge2.0^{+0.8}_{-0.6}$. Taking $k=\knr$ moves these to $10.2$ and $2.4$, a $13$ and $21\%$ shift, so the momentum parameter is not the dominant uncertainty in either case. The curvature ratio is. Writing $\lam_{\rm crit}$ for the value at which a curve first enters the $\Uone$ band $\tc=0.23$--$0.57$, the two reach it at $\lam_{\rm crit}=0.33$--$0.35$ and $0.62$--$0.68$, while the floor of Eq.~\eqref{eq:floor} becomes vacuous below $\lam=\Amp^{-1/2}$, that is $0.32$ and $0.58$. The same degeneracy read as an energy budget: at $\lam=k=1$ removal at contact supplies $90\%$ and $66\%$ of the coarsening rate, while the two channels contribute equally at $\lam=k\sqrt{2/\Amp}$, that is at $\lam=0.45$ and $0.82$ when $k=1$\footnote{Both follow from Sec.~\ref{sec:kbound}. Equation~\eqref{eq:floor} reads $\tc\ge\Amp\lam-1/\lam$, whose right-hand side changes sign at $\Amp\lam^{2}=1$; below $\lam=\Amp^{-1/2}$ it is negative and the bound is weaker than $\tc\ge0$. The fraction of $d(L^{2})/dt$ supplied by the sink is $\tc\kap/\Amp=1-k^{2}/(\Amp\lam^{2})$, which is $1-1/\Amp$ at $\lam=k=1$ and equals $\tfrac12$ when $\lam=k\sqrt{2/\Amp}$.}.

The statement the data support is therefore conditional on $\lam$, and its strength depends on which group is used. At $\lam=1$ the three thick-cell quenches place $\tc$ an order of magnitude above the oriented relativistic value, and the $\Delta P=2.00$~MPa quench a factor of a few above it; the excess disappears only if the curvature radius is smaller than the interstring spacing by a factor of about three in the first case and about $1.5$ in the second. We know of no measurement of $\lam$ in either system, and do not assume one. We also repeat the caution of Sec.~\ref{sec:whylc}: $\tc$ sums the two sink channels, so an elevated value is consistent with a $\Zt$ annihilation channel without singling it out.

\section{Discussion and outlook}\label{sec:disc}

The values $\tc=0.23$ and $0.57$~\cite{Martins2016}, and the more recent high-resolution field theory re-calibrations of the same order of magnitude~\cite{Correia:2019bdl,Correia:2021tok}, all describe relativistic, frictionless, oriented strings. We are not aware of a VOS calibration for a cosmological $\Zt$ network, so there is nothing to compare our result against at fixed topology, and the excess cannot be attributed to the $\Zt$ classification of the defects. There are concrete reasons to expect the one-scale hypothesis to behave better in cosmology than in a nematic. A local gauge $\Zt$ string has constant energy per unit length, with no logarithmic divergence to generate a second scale, and no long-range interaction of the kind disclinations have~\cite{ChuangPRE1993}. Nor does it have a type-1 analogue: the trivial class of a gauge vacuum manifold is the vacuum itself, so there is no contractible line that nonetheless persists, whereas in a nematic the type-1 lines are long-lived and are consumed by the type-$\frac{1}{2}$ network. This argues for simulating a friction-regime $\Zt$ network and measuring $\Amp$ directly, not for transporting our numbers.

The four available quenches give two amplitudes, $\Amp_{2}=10.0^{+1.3}_{-1.1}$ from the three in the $234~\mu$m cell and $\Amp_{1}=3.0^{+0.8}_{-0.6}$ from the one in the $158~\mu$m cell. Both are measured through a ratio that is dimensionless and free of the material constants of 5CB, because network coarsening and isolated-loop collapse are published for the same runs: the network obeys $L^{2}=\Amp\ld t$, an isolated loop obeys $r^{2}=2\ld(t_{0}-t)$, and the shared $\ld=K/\gamma$ cancels between them. The quoted intervals are restatements of the loop-collapse prior rather than measurements of scatter. Since $\Amp$ and $\ld$ appear only as the product $\Amp\ld$, the fit can never determine the amplitude better than the transport coefficient is known. Each quench contributes at most $1/s_{\ell}^{2}$ to the sum in Eq.~\eqref{eq:sigma}, however many points it carries, so $N_q$ quenches cannot give $\sigma_{\ln\Amp}$ below $s_{\ell}/\sqrt{N_q}$. The three thick-cell runs already supply $98\%$ of that maximum, giving $\sigma_{\ln\Amp}=0.116$ against a limit of $0.115$: more density points cannot help, and only more quenches with an independently known $\ld$ can.

That $k$, $\lam$ and $\tc$ enter only as $\Amp$ is a property of the overdamped attractor and not a limitation of the fit; no analysis of $\rho(t)$ can evade it. The two remaining freedoms are very different in size. The momentum parameter is bounded above by $k\le1$, and although nothing bounds it below, evaluating at $\knr=0.900$ instead moves $\tc$ by only $13$--$21\%$, so the range of plausible $k$ is not what limits the result. The curvature ratio is bounded neither way. At fixed $\Amp$, Eq.~\eqref{eq:ctilde} sends $\tc$ to zero as $\lam$ decreases, so any sink coefficient at all is compatible with the data for a sufficiently wiggly tangle. This is why Fig.~\ref{fig:lambda} closes in neither direction, and why $\lam$ dominates the uncertainty. It also controls the bound, since Eq.~\eqref{eq:floor} follows from $k\le1$, which converts to $\kap\le1$ only when $\lam\ge1$: a tangle appreciably wigglier than one-scale would allow $\kap>1$ and weaken the floor. What can be said without assuming $\lam$ is that both floors at $\lam=1$ exceed the $\Uone$ values, and that the weaker of the two still requires $\lam\lesssim0.68$ before the excess disappears.

We report both amplitudes without averaging them. A factor of about four between two determinations of the same quantity is the largest single feature of this analysis.  Within the thin-cell run, $\Aobs$ stays between $2.5$ and $2.9$ while $t$ runs over a factor four and the confinement ratio $L/d$---the correlation length as a fraction of the cell depth---over a factor two. It is not a difference of exponent either: $\nu_{1}=0.954$ and $\nu_{2}=1.030$ differ by too little to accumulate more than about $10\%$ across the observation window. Whatever causes the gap therefore acts as a constant multiplicative offset attached to the run. Appendix~\ref{app:matched} makes this quantitative: pairing thin-cell with thick-cell points that share the same value of $L/d$, $t$, $\rho$ or $\rho d$ leaves the ratio between them at $3.8$--$4.2$ against an expected value of $1$, with scatter no worse than the dispersion already present within the thick-cell group.

Several candidates are excluded immediately. A neglected initial condition gives $\Aobs=\Amp+L_{0}^{2}/(\ld t)$, which decreases monotonically with time, whereas the retained thin-cell values are not monotonic, running $2.89$, $2.77$, $2.58$, $2.48$, $2.54$, $2.60$ and $2.74$ in time order. The term is in any case positive, so subtracting it lowers rather than raises the inferred amplitude: fitting the two-parameter form to all points of each run gives $\Amp_{1}=2.14$ against $\Amp_{2}=7.3$--$11.8$, widening the separation. An error in the time origin would require a shift growing in proportion to $t$ itself---of order a few seconds at the earliest thick-cell point and tens of seconds at the latest---so no single value works, and any single value would visibly curve $\Aobs$. Anisotropy in the conversion of assumption (iv) is bounded from below. Writing $\Lambda=\xi\rho$, the factor is $\xi=1$ for lines running normal to the imaging plates, $\xi=2$ for an isotropic tangle, and $\xi>2$ for lines preferentially lying in the plane. Since $\Amp\propto1/\xi$, closing a factor $3.95$ would require $\xi_{1}\simeq0.5$, below the kinematic floor---and confinement-induced flattening, the effect one intuitively expects in a thin cell, raises $\xi$ and widens the gap rather than closing it. A gross calibration error also fails, at least in the direction usually assumed. A factor of two in linear scale would give exactly four in areal density, and would be constant per run. But dividing the thin-cell densities by four implies $L=233~\mu$m at the last retained point, against a cell depth of $158~\mu$m. We note, however, that the mirror hypothesis is not excluded by this test: nothing here establishes that it is $\Amp_{1}$ rather than $\Amp_{2}$ that is anomalous.

Three candidates remain, and they differ in what each would imply. The first is a genuine difference in tangle shape. With $\Amp\propto\lam^{-2}$ in the friction-dominated limit and $\propto\lam^{-1}$ in the sink-dominated one, two runs sharing a common $(k,\tc)$ require the ratio $\lam_{1}/\lam_{2}$ to lie between $(\Amp_{2}/\Amp_{1})^{1/2}$ and $\Amp_{2}/\Amp_{1}$. Using the $\nu\equiv1$ amplitudes, which Appendix~\ref{app:matched} identifies as the appropriate measure of a constant offset, this is $2.0$--$4.0$; using the fully free fits it is $1.8$--$3.4$. The $\Uone$ values for $\tc=0.23$--$0.57$ fall in the narrower range $1.9$--$2.0$, which matches the ratio of the two $\lam_{\rm crit}$ of Fig.~\ref{fig:lambda}. The difficulty with this explanation is that if confinement causes the shape difference, $\lam$ should track $L/d$, which doubles within the run while $\Aobs$ does not move. The mechanism would instead have to be memory of the formation stage frozen in at the quench, which the model does not derive.

The second is a difference in the momentum parameter. Section~\ref{sec:model} treats $k$ as a property of the dynamics rather than of the apparatus, but $k$ measures the alignment between velocity and curvature, and anchoring of the director at the cell surfaces is precisely an apparatus effect on alignment. Holding $\lam=1$ and $\tc=9.02$ fixed, the value $k_{1}=0.32$ reproduces $\Amp_{1}=2.98$ exactly. This fits as well as the shape hypothesis, and it has a sharper consequence: unlike a difference in $\lam$, it would undercut the $\kap\le1$ premise behind Eq.~\eqref{eq:floor}.

The third is a failure of assumption (v) in the thinner cell. If a significant fraction of lines terminate on the confining plates, they experience drag that the bulk loop-collapse measurement does not see, so the effective $\ld$ of the network would be smaller than the value assigned to it. We note that this is the one candidate the matched-pair test cannot address, since a wall term set by the areal density of line ends would be constant in $L/d$ by construction, and could not be removed by pairing points at equal $L/d$. We offer this as a possibility to be
tested rather than as an estimate; quantifying it would require knowing the fraction of wall-terminated lines, which the published data do not report. None of the three can be settled here. Reference~\cite{Chuang1991PRL} recorded ten pressure jumps at each $\Delta P$ and obtained its plotted errors by averaging over several runs, so $\Amp_{1}$ is not a single realization. What we
cannot do is test it against run-to-run scatter at fixed cell depth: the three thick-cell runs span $8.36$--$11.14$ with offsets $\{+29,-15,-9\}\%$, so a factor
four lies well outside their dispersion, but there is no second thin-cell run to confirm that the same holds there.

A number of limitations constrain the current analysis, each now accessible via standard three-dimensional confocal tracking of individual disclination lines~\cite{ZushiTakeuchi2022,ZushiSchimming2024}. These are ordered by their relative impact below. \emph{First, $\lam$.} It should be measured directly and separately for each quench, from the distribution of local curvature along tracked lines together with the interesting spacing obtained from the same frames, with a demonstration that
the ratio is constant over the observation window. This is the entire uncertainty in Fig.~\ref{fig:lambda}, and it is also what would show whether the two amplitudes reported here reflect two different tangle geometries. \emph{Second, $T/\Gamma$.} It should be resolved in pressure and temperature. Because $\Amp$ enters only as the product $\Amp\ld$, the unmatched published range \emph{is} the error bar. A loop-collapse measurement at the pressure and temperature of each quench, taken from the same recordings, would replace both Eq.~\eqref{eq:interp} and the prior width, and would incidentally resolve the temperature discrepancy between the source papers. \emph{Third, $k$.} It is in principle measurable: the correlation between velocity and curvature along tracked lines is exactly the average that defines it. This would also test whether $k$ differs between cells.

Beyond these, the line density $\Lambda$ should be measured from a three-dimensional reconstruction rather than inferred from planar crossings under an isotropy assumption, and all quenches should be run in a single cell thick enough that $L\ll d$ throughout---which would settle directly whether the amplitude difference reported here is a property of the cell. Finally, tracking that resolves individual reconnection and annihilation events would separate the two contributions to the sink, $\tceff=\tc_{\rm loop}+\tc_{\rm ann}$: the question that motivated this measurement, and the one the amplitude alone cannot answer.

\begin{table}[htbp]
  \centering
  \caption{String densities digitized from Fig.~2 of Ref.~\cite{Chuang1991PRL}, with the per-quench reading error $f_{j}$. Points in parentheses are omitted from all fits: the first and last of each run, following that paper, and additionally the second point of the $\Delta P=2.00$~MPa run, which begins later than the others. All four runs are used, the first as group~1 and the remaining three as
  group~2.}
  \label{tab:data}
  \begin{ruledtabular}
  \begin{tabular}{lcc}
    Run & $t$ (s) & $\rho$ (mm$^{-2}$) \\
    \hline
    $\Delta P=2.00$ & $(3.950)$ & $(148.02)$ \\
    $P_{f}=5.60$    & $(4.917)$ & $(142.71)$ \\
    $f_{j}=0.050$   & $6.292$   & $137.61$ \\
    $\ld=200$       & $7.907$   & $114.17$ \\
    $d=158~\mu$m    & $9.846$   & $98.35$ \\
                    & $12.488$  & $80.59$ \\
                    & $15.695$  & $62.82$ \\
                    & $19.908$  & $48.35$ \\
                    & $24.796$  & $36.75$ \\
                    & $(31.736)$& $(30.50)$ \\
    \hline
    $\Delta P=2.28$ & $(1.975)$ & $(109.28)$ \\
    $P_{f}=5.88$    & $3.355$   & $82.15$ \\
    $f_{j}=0.056$   & $5.700$   & $55.17$ \\
    $\ld=210$       & $9.690$   & $28.49$ \\
    $d=234~\mu$m    & $16.466$  & $17.75$ \\
                    & $(28.235)$& $(11.49)$ \\
    \hline
    $\Delta P=2.62$ & $(0.995)$ & $(150.83)$ \\
    $P_{f}=6.22$    & $1.488$   & $125.25$ \\
    $f_{j}=0.056$   & $2.266$   & $86.21$ \\
    $\ld=223$       & $3.357$   & $58.60$ \\
    $d=234~\mu$m    & $5.020$   & $35.59$ \\
                    & $7.508$   & $21.88$ \\
                    & $(11.224)$& $(16.65)$ \\
    \hline
    $\Delta P=4.69$ & $(0.996)$ & $(129.80)$ \\
    $P_{f}=8.29$    & $1.298$   & $111.84$ \\
    $f_{j}=0.056$   & $1.676$   & $91.65$ \\
    $\ld=300$       & $2.185$   & $68.81$ \\
    $d=234~\mu$m    & $2.875$   & $51.03$ \\
                    & $3.647$   & $37.83$ \\
                    & $4.754$   & $30.24$ \\
                    & $6.198$   & $23.87$ \\
                    & $(8.152)$ & $(19.56)$ \\
  \end{tabular}
  \end{ruledtabular}
\end{table}

\begin{acknowledgments}
We warmly thank Nikolaos Lattas for useful discussions, his assistance with data digitization and for introducing the ideas behind Appendix B. This research was supported by COST Action CA21136 - Addressing observational tensions in cosmology with systematics and fundamental physics (CosmoVerse), supported by COST (European Cooperation in Science and Technology).
\end{acknowledgments}

\section*{Data availability}
The digitized data of Table~\ref{tab:data}, the likelihood of Eq.~\eqref{eq:chi2} and the code producing all figures constitute the whole of the analysis, and are available as a CSV dataset and Mathematica (v13) notebook deposited on GitHub~\cite{efstratiou_github_2026}.

\appendix
\section{Data digitization and point selection}
\label{app:data}

\subsection{Extraction}

The string density data from Fig.~2 of Ref.~\cite{Chuang1991PRL} were extracted using the WebPlotDigitizer tool. The plot axes were first calibrated on a log-log scale, using the major tick marks of the original figure as reference points. Following calibration, independent datasets were created for each distinct marker shape (plus, triangle, asterisk and diamond), which correspond to the four pressure quenches $\Delta P=2.00$, $2.28$, $2.62$ and $4.69$~MPa. The time and density coordinates $(t,\rho)$ of each point were then extracted manually into distinct sets.

We follow Ref.~\cite{Chuang1991PRL} in omitting the first and last point of each run. That paper attributes the departures from a straight line at the two ends of each data set to the finite camera resolution, which thickens the line images so that overlapping strings are undercounted at early times, and to image noise, which inflates the count at late times. We drop the second point of the $\Delta P=2.00$~MPa run as well, since it lies $26\%$ below the $\nu=1$ line of Fig.~2 of Ref.~\cite{Chuang1991PRL}, against $5$, $1$ and $1\%$ for the second points of the other three runs. All retained points lie within $16<\rho<160~\mathrm{mm}^{-2}$, the range over which Ref.~\cite{Chuang1991PRL} quotes its exponent. The selection does not produce the separation between the two groups: retaining the second point gives $\Amp_{1}=2.75$ rather than $2.65$, and retaining every point of every run gives single-quench amplitudes of $2.86$, $8.48$, $12.52$ and $11.43$, against the $2.65$, $8.23$, $12.19$ and $11.38$ obtained from the retained points alone. In every case the shift is smaller than the prior width, and the two groups remain separated by the same factor.

\subsection{Reading errors}

Because the original data lacked explicit error bars, the reading uncertainties were estimated from the physical size of the plotted symbols. We assumed that the fractional half-height of a symbol corresponds to a $1\sigma$ uncertainty in logarithmic space. Digitizing the upper ($y_{\rm top}$) and lower ($y_{\rm bottom}$) bounds of a representative symbol for each dataset, the fractional error is
\begin{equation}
  f_j\equiv\sigma_{\ln\rho}\approx\frac{y_{\rm top}-y_{\rm bottom}}{y_{\rm top}+y_{\rm bottom}}.
\end{equation}
This yielded a uniform relative uncertainty per dataset---$0.050$ for the $158~\mu$m cell run and $0.056$ for the $234~\mu$m cell runs---applied to every point of the corresponding quench. The complete digitized datasets are summarized in Table~\ref{tab:data}.

\section{Testing whether the gap tracks any measured property of the runs}
\label{app:matched}

\subsection{The construction}

Suppose the observed amplitude were the true amplitude multiplied by an unknown, strictly positive distortion factor $B(X)$ that depends on some measured property $X$ of the data point. That property could be the confinement ratio $L/d$, the elapsed time $t$, the density $\rho$, or the optical crowding $\rho d$.

The last of these deserves a word, since it is the covariate built specifically to capture occlusion. Because $\Lambda=2\rho$ is the length of line per unit volume, $2\rho d$ is the total length of line held in a column of unit cross-section spanning the cell, that is, the length of line per unit area of the projected image. Equivalently, $\rho d$ is the number of line images crossing unit length of any test line drawn in that image; it carries dimensions of inverse length, and only ratios of it enter below. Multiplied by the apparent width of a line it gives the fraction of the field of view covered by defect images, and so it is the quantity that decides whether two lines overlap and are counted as one. Matching on $\rho d$ is therefore not the same test as matching on $\rho$: at equal $\rho$ the thinner cell holds less line along the viewing direction and its image is correspondingly less crowded, so the two rows probe occlusion depending on the areal density and on the projected line content respectively.

The test exploits a cancellation. Take a data point from the thin cell and pair it with one from the thick cell at the same value of $X$. The unknown factor $B(X)$
is then identical for the two, and cancels exactly in their ratio $q\equiv\Aobs^{\rm thick}/\Aobs^{\rm thin}$. The virtue of this construction is that we never need to know what $B$ is; we need only assume that it depends on $X$ alone. If the two runs share the same underlying amplitude, the paired ratio must come out at $q=1$.

Two conditions are needed for a pair to exist at all, and neither is automatic. First, two independently sampled runs never share an exact value of $X$, so we match within a tolerance, keeping cross-cell pairs from the retained points of Table~\ref{tab:data} with $|\ln(X_{1}/X_{2})|\le\epsilon=0.06$. Second, a thin-cell point has a partner only if its value of $X$ lies inside the range the thick-cell runs span. Where it does not, the point is dropped and the test says nothing about it, so each row of Table~\ref{tab:matched_test} interrogates only the part of the thin-cell run that the thick-cell runs reach. The Support column records that fraction. For $\rho d$ the thin-cell range lies wholly inside the thick-cell one and the coverage is complete; for $L/d$ and $\rho$ it is $95.6\%$ and $92.9\%$. For $t$ it is only $70.1\%$: the thick-cell runs stop at $16.5$~s while the thin-cell run continues to $24.8$~s, so the $t$ matching tests the first two-thirds of that run and is blind to its late-time tail, which is where confinement would be most severe. This, as much as the small number of pairs, is why we call the $t$ row suggestive rather than conclusive.

One further condition is implicit in the cancellation: $B$ must be the \emph{same} function of $X$ in both cells. For $L/d$ and $\rho d$, which already contain the cell depth, that is the natural reading of the hypothesis being tested. For $t$ and $\rho$ it is an additional restriction, since a distortion depending on elapsed time and cell depth separately would not cancel between the two members of a pair.

Throughout we use the raw, point-by-point amplitudes assuming $\nu\equiv1$, stripped of any overall fit offsets $\hat\delta_{j}$.

\subsection{Result}

As Table~\ref{tab:matched_test} shows, the paired ratio stays at $3.8$--$4.2$ no matter which property is matched. Given that the raw averages of the two groups already differ by a factor $4.04$, pairing point by point removes essentially none of the discrepancy. The scatter among the pairs, measured as the standard deviation of $\ln q$, is $0.11$--$0.20$, no worse than the dispersion $0.169$ of $\ln\Aobs$ among the sixteen retained thick-cell points themselves.

The size of the effect is best stated relative to that scatter. The measured offsets are $\ln\langle q\rangle=1.33$--$1.43$ against a null of zero, so the gap is $7$ to $13$ times the scatter of a single pair. We quote it this way, rather than as a standard error on the mean, because the pairs are not independent: only seven thin-cell points survive the selection of Table~\ref{tab:data}, so the twelve pairs formed at matched $L/d$ necessarily reuse the same thin-cell points against several thick-cell partners. $N$ in Table~\ref{tab:matched_test} is therefore a count of comparisons and not of independent measurements, and dividing ${\rm sd}(\ln q)$ by $\sqrt{N}$ would overstate the significance. Nothing in the argument needs it: the offset exceeds the per-pair scatter by so wide a margin that no plausible correlation among the pairs would bring $q$ near unity.
\begin{table}[htbp]
  \centering
  \caption{Matched-pair test at $\epsilon=0.06$. Each row pairs thin-cell with thick-cell data points that share a common value of the property $X$, so that any distortion depending on $X$ alone cancels in the ratio $q\equiv\Aobs^{\rm thick}/\Aobs^{\rm thin}$. $N$ is the number of such pairs; because only seven thin-cell points are retained, the pairs reuse points and $N$ counts comparisons rather than independent measurements. $\langle q\rangle$ is their geometric mean, the exponential of the mean of $\ln q$, whose expected value is $1$ if the two groups share an amplitude; the observed $\ln\langle q\rangle=1.33$--$1.43$ therefore exceeds the per-pair scatter ${\rm sd}(\ln q)$ by a factor of $7$ to $13$. That scatter is in turn no larger than $0.169$, the dispersion of $\ln\Aobs$ among the sixteen retained thick-cell points. Support is the fraction of the thin-cell run's logarithmic range in $X$ that the thick-cell runs cover, and so measures how much of that run the test actually probes; the matching in $t$, at $70.1\%$, is the one row that does not cover most of it.}
  \label{tab:matched_test}
  \begin{ruledtabular}
  \begin{tabular}{lcccc}
    Property $X$ & $N$ & $\langle q\rangle$ & $\mathrm{sd}(\ln q)$ & Support \\
    \hline
    $L/d$    & $12$ & $4.03$ & $0.203$ & $95.6\%$ \\
    $t$      & $4$  & $3.79$ & $0.194$ & $70.1\%$ \\
    $\rho$   & $5$  & $4.18$ & $0.108$ & $92.9\%$ \\
    $\rho d$ & $4$  & $3.77$ & $0.152$ & $100\%$ \\
  \end{tabular}
  \end{ruledtabular}
\end{table}
The four named mechanisms are therefore excluded: the gap is not a smooth function of boundary confinement or orientation flattening (which would act through $L/d$), of residual initial conditions or of network drift toward the midplane (through $t$), or of optical occlusion (through $\rho$ and $\rho d$). It is worth singling out the last of these: $\rho d$ is both the covariate constructed to measure crowding and the only one whose thin-cell range lies wholly inside the thick-cell range, so it is tested over the entire run, and it still returns $3.77$. Occlusion is the systematic this test excludes most cleanly. At the other extreme, the matching in $t$ yields only four pairs and covers only the first two-thirds of the thin-cell run, and is suggestive rather than conclusive.

\subsection{What the test does not show}

The test excludes distortions that \emph{vary} with the matched property. A distortion that is constant within a run passes it invisibly, because $B$ cancels whether or not it depends on anything at all. The correct conclusion is therefore the weaker one: the offset is not a smooth function of confinement, elapsed time or crowding, and behaves instead as a constant attached to the run. Which constant, the test does not say---and each of the candidates that survive in Sec.~\ref{sec:disc}, namely a $\lam$ frozen in at formation, a difference in $k$, and a wall-drag contribution set by the areal density of line ends, has exactly the shape the test is blind to.

The test is also silent on \emph{which} group is anomalous, since it measures only a ratio. A systematic residing in the thick-cell runs would look identical.

Two consequences for the main text follow. First, since the offset is a constant per run, the appropriate measure of the gap is $3.95=10.46/2.65$, taken from the $\nu\equiv1$ rows of Table~\ref{tab:fits}, rather than the $10.02/2.98$ ratio of the fully free fits: a constant offset should not be partially absorbed into a free exponent. Second, this justifies fitting the two groups separately rather than averaging them.

\subsection{Partially separating cell depth from quench depth}

The $\Delta P=2.00$~MPa run is ambiguous in its interpretation, being both the shallowest quench and the only one in the thin cell. The two can be partly separated using group~2 alone, where $\Delta P$ spans a factor of two while $\Amp$ moves only from $8.36$ to $11.14$. The local slope between the two shallowest thick-cell runs is $\simeq7.8$ per MPa, which extrapolated down to $\Delta P=2.00$ predicts $\Amp\simeq6$---still twice the observed $2.98$, and with a trend of the wrong sign to begin with, since within group~2 the amplitude \emph{increases} with quench depth. Quench depth alone would therefore have to behave very differently below $2.28$~MPa to produce the observed gap. This points toward the cell, or toward something specific to that one recording, rather than toward quench depth.

\bibliography{new_refs}

@article{Kibble1976,
  author  = {Kibble, T. W. B.},
  title   = {Topology of cosmic domains and strings},
  journal = {J. Phys. A: Math. Gen.},
  volume  = {9},
  pages   = {1387--1398},
  year    = {1976},
  doi     = {10.1088/0305-4470/9/8/029}
}

@article{Kibble1980,
  author  = {Kibble, T. W. B.},
  title   = {Some implications of a cosmological phase transition},
  journal = {Phys. Rep.},
  volume  = {67},
  pages   = {183},
  year    = {1980},
  doi     = {10.1016/0370-1573(80)90091-5}
}

@book{VilenkinShellard,
  author    = {Vilenkin, A. and Shellard, E. P. S.},
  title     = {Cosmic Strings and Other Topological Defects},
  publisher = {Cambridge University Press},
  address   = {Cambridge},
  year      = {2000},
  isbn      = {9780521654760}
}

@book{nakahara2003geometry,
  author    = {Nakahara, Mikio},
  title     = {Geometry, Topology and Physics},
  edition   = {2nd},
  publisher = {CRC Press},
  address   = {Boca Raton},
  year      = {2003}
}

@article{Zurek1985,
  author  = {Zurek, W. H.},
  title   = {Cosmological experiments in superfluid helium?},
  journal = {Nature},
  volume  = {317},
  pages   = {505},
  year    = {1985},
  doi     = {10.1038/317505a0}
}

@article{Zurek1996,
  author  = {Zurek, W. H.},
  title   = {Cosmological experiments in condensed matter systems},
  journal = {Phys. Rep.},
  volume  = {276},
  pages   = {177},
  year    = {1996},
  doi     = {10.1016/S0370-1573(96)00009-9}
}

@article{MartinsShellard1996,
  author  = {Martins, C. J. A. P. and Shellard, E. P. S.},
  title   = {Scale-invariant string evolution with friction},
  journal = {Phys. Rev. D},
  volume  = {53},
  pages   = {R575},
  year    = {1996},
  doi     = {10.1103/PhysRevD.53.R575},
  eprint  = {hep-ph/9507335},
  archivePrefix = {arXiv}
}

@article{MartinsShellard2002,
  author  = {Martins, C. J. A. P. and Shellard, E. P. S.},
  title   = {Extending the velocity-dependent one-scale string evolution model},
  journal = {Phys. Rev. D},
  volume  = {65},
  pages   = {043514},
  year    = {2002},
  doi     = {10.1103/PhysRevD.65.043514},
  eprint  = {hep-ph/0003298},
  archivePrefix = {arXiv}
}

@article{MooreMartinsShellard2002,
  author  = {Moore, J. N. and Shellard, E. P. S. and Martins, C. J. A. P.},
  title   = {Evolution of {A}belian-{H}iggs string networks},
  journal = {Phys. Rev. D},
  volume  = {65},
  pages   = {023503},
  year    = {2002},
  doi     = {10.1103/PhysRevD.65.023503},
  eprint  = {hep-ph/0107171},
  archivePrefix = {arXiv}
}

@article{MartinsMooreShellard2004,
  author  = {Martins, C. J. A. P. and Moore, J. N. and Shellard, E. P. S.},
  title   = {Unified model for vortex-string network evolution},
  journal = {Phys. Rev. Lett.},
  volume  = {92},
  pages   = {251601},
  year    = {2004},
  doi     = {10.1103/PhysRevLett.92.251601},
  eprint  = {hep-ph/0310255},
  archivePrefix = {arXiv}
}

@book{Martins2016,
  author    = {Martins, C. J. A. P.},
  title     = {Defect Evolution in Cosmology and Condensed Matter: Quantitative Analysis with the Velocity-Dependent One-Scale Model},
  series    = {SpringerBriefs in Physics},
  publisher = {Springer},
  address   = {Cham},
  year      = {2016},
  doi       = {10.1007/978-3-319-44553-3}
}

@article{Vanchurin2005,
  author  = {Vanchurin, V. and Olum, K. D. and Vilenkin, A.},
  title   = {Cosmic string scaling in flat space},
  journal = {Phys. Rev. D},
  volume  = {72},
  pages   = {063514},
  year    = {2005},
  doi     = {10.1103/PhysRevD.72.063514},
  eprint  = {gr-qc/0501040},
  archivePrefix = {arXiv}
}

@article{BlancoPilladoOlumShlaer2014,
  author  = {Blanco-Pillado, J. J. and Olum, K. D. and Shlaer, B.},
  title   = {The number of cosmic string loops},
  journal = {Phys. Rev. D},
  volume  = {89},
  pages   = {023512},
  year    = {2014},
  doi     = {10.1103/PhysRevD.89.023512},
  eprint  = {1309.6637},
  archivePrefix = {arXiv}
}

@article{Auclair2020,
  author  = {Auclair, P. and Blanco-Pillado, J. J. and Figueroa, D. G.
             and Jenkins, A. C. and Lewicki, M. and Sakellariadou, M.
             and Sanidas, S. and Sousa, L. and Steer, D. A.
             and Wachter, J. M. and Kuroyanagi, S.},
  title   = {Probing the gravitational wave background from cosmic strings with {LISA}},
  journal = {JCAP},
  volume  = {2020},
  number  = {04},
  pages   = {034},
  year    = {2020},
  doi     = {10.1088/1475-7516/2020/04/034},
  eprint  = {1909.00819},
  archivePrefix = {arXiv}
}

@article{NANOGrav2023,
  author  = {{NANOGrav Collaboration} and Afzal, A. and others},
  title   = {The {NANOGrav} 15 yr data set: Search for signals from new physics},
  journal = {Astrophys. J. Lett.},
  volume  = {951},
  pages   = {L11},
  year    = {2023},
  doi     = {10.3847/2041-8213/acdc91},
  eprint  = {2306.16219},
  archivePrefix = {arXiv}
}

@article{Chuang1991PRL,
  author  = {Chuang, I. and Turok, N. and Yurke, B.},
  title   = {Late-time coarsening dynamics in a nematic liquid crystal},
  journal = {Phys. Rev. Lett.},
  volume  = {66},
  pages   = {2472--2475},
  year    = {1991},
  doi     = {10.1103/PhysRevLett.66.2472}
}

@article{Chuang1991Science,
  author  = {Chuang, I. and Durrer, R. and Turok, N. and Yurke, B.},
  title   = {Cosmology in the laboratory: Defect dynamics in liquid crystals},
  journal = {Science},
  volume  = {251},
  number  = {4999},
  pages   = {1336--1342},
  year    = {1991},
  doi     = {10.1126/science.251.4999.1336}
}

@article{ChuangPRE1993,
  author  = {Chuang, I. and Yurke, B. and Pargellis, A. N. and Turok, N.},
  title   = {Coarsening dynamics in uniaxial nematic liquid crystals},
  journal = {Phys. Rev. E},
  volume  = {47},
  pages   = {3343--3356},
  year    = {1993},
  doi     = {10.1103/PhysRevE.47.3343}
}

@article{BowickChandar1994,
  author  = {Bowick, M. J. and Chandar, L. and Schiff, E. A. and Srivastava, A. M.},
  title   = {The cosmological {K}ibble mechanism in the laboratory: String formation in liquid crystals},
  journal = {Science},
  volume  = {263},
  number  = {5149},
  pages   = {943--945},
  year    = {1994},
  doi     = {10.1126/science.263.5149.943}
}

@article{Pargellis1992,
  author  = {Pargellis, A. N. and Finn, P. and Goodby, J. W. and Panizza, P. and Yurke, B. and Cladis, P. E.},
  title   = {Defect dynamics and coarsening dynamics in smectic-$C$ films},
  journal = {Phys. Rev. A},
  volume  = {46},
  pages   = {7765},
  year    = {1992},
  doi     = {10.1103/PhysRevA.46.7765}
}

@article{ZushiTakeuchi2022,
  author  = {Zushi, Y. and Takeuchi, K. A.},
  title   = {Scaling and spontaneous symmetry restoring of topological defect dynamics in liquid crystal},
  journal = {Proc. Natl. Acad. Sci. U.S.A.},
  volume  = {119},
  pages   = {e2207349119},
  year    = {2022},
  doi     = {10.1073/pnas.2207349119},
  eprint  = {2110.00442},
  archivePrefix = {arXiv}
}

@misc{ZushiSchimming2024,
  author        = {Zushi, Y. and Schimming, C. D. and Takeuchi, K. A.},
  title         = {Approach and rotation of reconnecting topological defect lines in liquid crystal},
  year          = {2024},
  eprint        = {2404.01480},
  archivePrefix = {arXiv},
  primaryClass  = {cond-mat.soft}
}

@article{ImuraOkano,
  author  = {Imura, H. and Okano, K.},
  title   = {Friction coefficient for a moving disinclination in a nematic liquid crystal},
  journal = {Phys. Lett. A},
  volume  = {42},
  number  = {5},
  pages   = {403--404},
  year    = {1973},
  doi     = {10.1016/0375-9601(73)90728-7}
}

@book{deGennesProst,
  author    = {de Gennes, P. G. and Prost, J.},
  title     = {The Physics of Liquid Crystals},
  edition   = {2nd},
  publisher = {Clarendon Press},
  address   = {Oxford},
  year      = {1993},
  isbn      = {9780198517856}
}

@book{KlemanLavrentovich,
  author    = {Kl{\'e}man, M. and Lavrentovich, O. D.},
  title     = {Soft Matter Physics: An Introduction},
  publisher = {Springer},
  address   = {New York},
  year      = {2003},
  doi       = {10.1007/b97416}
}

@article{Bray1994,
  author  = {Bray, A. J.},
  title   = {Theory of phase-ordering kinetics},
  journal = {Adv. Phys.},
  volume  = {43},
  pages   = {357},
  year    = {1994},
  doi     = {10.1080/00018739400101505}
}

@book{Underwood1970,
  author    = {Underwood, E. E.},
  title     = {Quantitative Stereology},
  publisher = {Addison-Wesley},
  address   = {Reading, MA},
  year      = {1970}
}

@article{Martins:1997nb,
    author = "Martins, C. J. A. P. and Shellard, E. P. S.",
    title = "{Evolution of superconducting string currents}",
    eprint = "hep-ph/9706533",
    archivePrefix = "arXiv",
    reportNumber = "DAMTP-R-97-29",
    doi = "10.1016/S0370-2693(98)00610-8",
    journal = "Phys. Lett. B",
    volume = "432",
    pages = "58--64",
    year = "1998"
}

@misc{efstratiou_github_2026,
author = {Efstratiou, Dimitrios and Paraskevas, Evangelos Achilleas and Perivolaropoulos,
, Leandros},
title = {Data and analysis notebook for the friction-era {VOS} amplitude of a {$\mathbb{Z}_2$} string network},
year = {2026},
howpublished = {GitHub repository, \url{https://github.com/Dimitrios1993/friction-era-vos-amplitude-z2-nlc}}
}

@article{PhysRevB.56.10892,
  title = {Averaged methods for vortex-string evolution},
  author = {Martins, C. J. A. P. and Shellard, E. P. S.},
  journal = {Phys. Rev. B},
  volume = {56},
  issue = {17},
  pages = {10892--10906},
  numpages = {0},
  year = {1997},
  month = {Nov},
  publisher = {American Physical Society},
  doi = {10.1103/PhysRevB.56.10892},
  url = {https://link.aps.org/doi/10.1103/PhysRevB.56.10892}
}

@article{PhysRevD.54.2535,
  title = {Quantitative string evolution},
  author = {Martins, C. J. A. P. and Shellard, E. P. S.},
  journal = {Phys. Rev. D},
  volume = {54},
  issue = {4},
  pages = {2535--2556},
  numpages = {0},
  year = {1996},
  month = {Aug},
  publisher = {American Physical Society},
  doi = {10.1103/PhysRevD.54.2535},
  url = {https://link.aps.org/doi/10.1103/PhysRevD.54.2535}
}

@article{Correia:2021tok,
    author = "Correia, J. R. C. C. C. and Martins, C. J. A. P.",
    title = "{High resolution calibration of the cosmic strings velocity dependent one-scale model}",
    eprint = "2108.07513",
    archivePrefix = "arXiv",
    primaryClass = "astro-ph.CO",
    doi = "10.1103/PhysRevD.104.063511",
    journal = "Phys. Rev. D",
    volume = "104",
    number = "6",
    pages = "063511",
    year = "2021"
}

@article{Correia:2019bdl,
    author = "Correia, J. R. C. C. C. and Martins, C. J. A. P.",
    title = "{Extending and Calibrating the Velocity dependent One-Scale model for Cosmic Strings with One Thousand Field Theory Simulations}",
    eprint = "1911.03163",
    archivePrefix = "arXiv",
    primaryClass = "astro-ph.CO",
    doi = "10.1103/PhysRevD.100.103517",
    journal = "Phys. Rev. D",
    volume = "100",
    number = "10",
    pages = "103517",
    year = "2019"
}

@article{Coelho:2026oeg,
    author = "Coelho, C. S. C. M. and Gschrey, A. -L. Y. and Martins, C. J. A. P.",
    title = "{Scaling solutions for varying tension strings}",
    eprint = "2602.18586",
    archivePrefix = "arXiv",
    primaryClass = "hep-ph",
    doi = "10.1103/k1ws-wsst",
    journal = "Phys. Rev. D",
    volume = "113",
    number = "6",
    pages = "063516",
    year = "2026"
}

@article{McGraw:1997nx,
    author = "McGraw, Patrick",
    title = "{Evolution of a nonAbelian cosmic string network}",
    eprint = "astro-ph/9706182",
    archivePrefix = "arXiv",
    doi = "10.1103/PhysRevD.57.3317",
    journal = "Phys. Rev. D",
    volume = "57",
    pages = "3317--3339",
    year = "1998"
}
\end{document}